\documentclass[]{jfm}

\usepackage{graphicx}
\usepackage{newtxtext}
\usepackage{newtxmath}
\usepackage{textcomp,mathcomp}
\usepackage{bm}
\usepackage{multirow}
\usepackage{natbib}
\usepackage{hyperref}

\hypersetup{
    colorlinks = true,
    urlcolor   = blue,
    citecolor  = black,
}

\newcommand{\RomanNumeralCaps}[1]
\linenumbers

\usepackage{float}

\usepackage{tabularx}

\title{Long-Pulse Laser-Induced Cavitation: A Race Between Advection and Phase Transition}

\author{Xuning Zhao\aff{1},
  Wentao Ma\aff{1},
  Junqin Chen\aff{2},
  Gaoming Xiang\aff{2,3,4},
  Pei Zhong\aff{2}
 \and Kevin Wang\aff{1}
 \corresp{\email{kevinw3@vt.edu}}}

\affiliation{\aff{1}Kevin T. Crofton Department of Aerospace and Ocean Engineering, Virginia Tech, Blacksburg, VA 24061, USA
\aff{2}Thomas Lord Department of Mechanical Engineering and Materials Science, Duke University, Durham, NC 27708, USA
\aff{3}Optics and Thermal Radiation Research Center, Institute of Frontier and Interdisciplinary Science, Shandong University, Qingdao, 266237, China
\aff{4}School of Energy and Power Engineering, Shandong University, Jinan, Shandong 250061, China}

\begin{document}
\maketitle

\begin{abstract}

Vapor bubbles generated by long-pulsed laser often have complex non-spherical shapes that reflect some characteristics (e.g., direction, width) of the laser beam. The transition between two commonly observed shapes --- namely, a rounded pear-like shape and an elongated conical shape --- is studied using a new computational model that combines compressible multiphase fluid dynamics with laser radiation and phase transition. Two laboratory experiments are simulated, in which Holmium:YAG and Thulium fiber lasers are used separately to generate bubbles of different shapes. In both cases, the bubble morphology predicted by the simulation agrees reasonably well with the experimental measurement. The simulated laser radiance, temperature, velocity, and pressure fields are analyzed to explain bubble dynamics and energy transmission. It is found that due to the lasting energy input (i.e.~long-pulsed laser), the vapor bubble's dynamics is driven not only by advection, but also by the continuation of vaporization. Notably, vaporization lasts less than $1$ microsecond in the case of the pear-shaped bubble, versus more than $50$ microseconds for the elongated bubble. It is hypothesized that the bubble's shape is the result of a competition. When the speed of advection is higher than that of vaporization, the bubble tends to grow spherically. Otherwise, it elongates along the laser beam direction. To clarify and test this hypothesis, the two speeds are defined analytically using a simplified model, then estimated for the experiments using simulation results. The results support the hypothesis. They also suggest that a higher laser absorption coefficient and a narrower beam facilitate bubble elongation.



\end{abstract}


\section{Introduction}
\label{sec:intro}

Vapor bubbles appear in many scientific studies and real-world applications that involve laser radiation. To researchers who study cavitation and bubble dynamics, laser is a convenient tool to create bubbles at a precise location without too much disturbance to the surrounding environment  \citep{zwaan2007controlled,tomita2002growth,brujan_nahen_schmidt_vogel_2001}. To technology developers and practitioners who use high-power laser in a liquid environment, cavitation is often an inevitable phenomenon, and the resulting vapor bubbles may have both beneficial and detrimental effects. Example applications in this regard include liquid-assisted laser processing (e.g., underwater laser cutting \citep{chida2003underwater}, laser cleaning \citep{ohl2006surface,song2004laser}), ocular laser surgery \citep{vogel1986cavitation,povzar2020cavitation}, laser angioplasty \citep{vogel1996minimization}, and laser lithotripsy \citep{fried2018advances,ho2021role}. The most common effects of laser-induced cavitation include the creation of a vapor channel, the disturbance of the local flow field, a propulsive force from bubble expansion, and material damages caused by the shock waves and micro-jets from bubble collapse \citep{chen2022cavitation,xiang2023dissimilar,dijkink2008laser,mohammadzadeh2015bubble}. Improving a technology often requires optimizing the trade-offs between these effects.

While laser-induced cavitation can be roughly described as localized boiling through thermal radiation, the detailed physics involves laser emission and absorption, phase transition, and the dynamics and thermodynamics of a two-phase fluid flow. Within this multiphysics problem, a key external (i.e.~user-specified) parameter is the duration of the laser pulse. In different applications, the value of this parameter varies from femtoseconds ($10^{-15}$ s) to more than one second \citep{zwaan2007controlled,ho2021role,juhasz1996time}. When the pulse duration is much smaller than the acoustic time scale in the fluid medium (i.e.,~characteristic length divided by sound speed), the laser energy input is referred to as short-pulsed. In this case, laser radiation can be assumed to be a preceding event that ends before the vapor bubble starts to expand. Therefore, the analysis of fluid and bubble dynamics can be separated from that of laser radiation, which simplifies the problem \citep{zein2013modeling,koch2016numerical,byun2004model}. In this paper, we study cavitation induced by long-pulsed laser, which means the duration of the laser pulse is comparable to or longer than the acoustic time scale. In this case, laser radiation may continue after the formation of the initial bubble. The assumption mentioned above is no longer valid. Laser radiation, phase transition, and the fluid dynamics and thermodynamics are now interdependent. They need to be analyzed together.

\begin{figure}
  \centerline{\includegraphics[width = 0.8\textwidth]{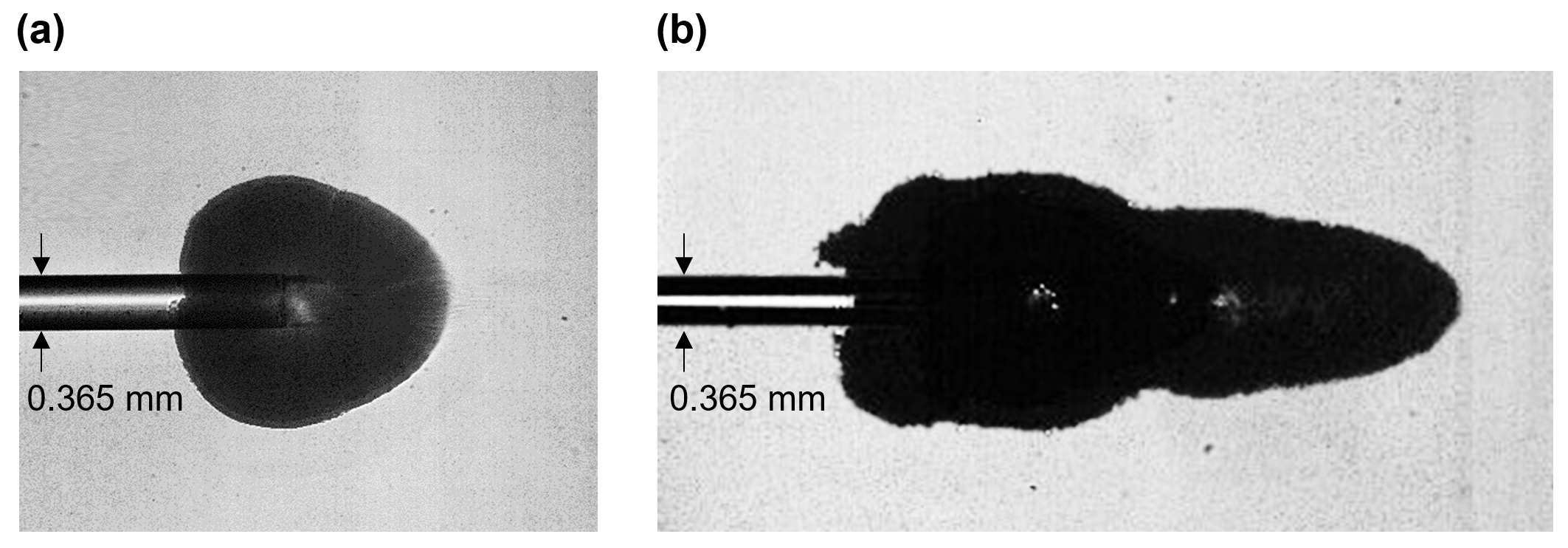}}
  \caption{Non-spherical vapor bubbles generated by long-pulsed laser. (a) A rounded, pear-shaped bubble generated by Ho:YAG laser (wavelength: $2080~\text{nm}$, pulse energy: $0.2~\text{J}$, pulse duration: $70~\text{\textmu s}$, acoustic time scale (fiber diameter divided by sound speed): $0.25~\text{\textmu s}$). (b) An elongated bubble generated by Thulium fiber laser (wavelength: $1940~\text{nm}$, pulse energy: $0.11~\text{J}$, pulse duration: $170~\text{\textmu s}$, acoustic time scale: $0.25~\text{\textmu s}$).}
 \label{fig:two_bubbles}
\end{figure}

The vapor bubbles generated by long-pulsed laser often have a non-spherical shape that reflects some characteristics of the laser beam, such as its direction and width. Figure~\ref{fig:two_bubbles} shows two commonly observed shapes that will be studied in this paper, namely a rounded pear-like shape and an elongated conical shape. Depending on the application, one or the other may be preferred. For example, a rounded bubble, when collapses, is more likely to generate a strong liquid jet that damages a surrounding material \citep{xiang2023dissimilar,chen2022cavitation}. An elongated bubble, on the other hand, can be more energy efficient in creating a long vapor channel that allows laser to pass through. While bubbles of both shapes have been observed in many experiments \citep{mohammadzadeh2015bubble,hardy2016cavitation,fried2018advances,xiang2023dissimilar}, the causal relation between bubble morphology and laser setting (e.g., wavelength, power magnitude and distribution, pulse duration, diverging angle) is still unclear. Many fundamental questions of practical significance are unresolved, such as the following.
\begin{itemize}
\item Does phase transition (vaporization) last for a substantial period of time, or does it occur instantaneously?
\item What fraction of the laser energy input is used to create the vapor bubble?
\item How to control the laser setting so that the vapor bubble has a desired shape?
\end{itemize}

It is difficult to answer these questions by laboratory experiment only. Some parameters of laser (e.g., wavelength) cannot be continuously varied in experiments. While the evolution of bubble shape can be measured by high-speed optical imaging, measurement of the pressure, velocity, and temperature fields inside and around the vapor bubble is challenging to say the least \citep{dular2013thermodynamic,petkovvsek2013ir,khlifa2013velocity}. Therefore, the partition of energy and the cause of the bubble's shape change cannot be easily determined. In this work, we combine laboratory experiment with numerical simulation to study long-pulse laser-induced cavitation, focusing on the physics behind pear-shaped and elongated bubbles. We try to investigate the causal relation between laser setting and the vapor bubble's shape, and to gain some insight on the three open questions mentioned above. 

A key novelty in this work is the use of a new computational model that combines compressible multiphase fluid dynamics with laser radiation and phase transition. In the past, bubble dynamics simulations were typically based on the solution of Rayleigh-Plesset, boundary integral, or multi-dimensional Navier-Stokes equations \citep{warnez2015numerical,klaseboer2006simulations,wang2017multiphase,cao2021shock}. Thermal radiation and the resulting phase transition were not included. A simulation usually starts with one or multiple spherical bubbles as the initial condition. In most cases, the initial state inside each bubble is set to a constant. This approach can be justified for bubbles generated by short-pulsed laser, given that radiation and vaporization both complete at a smaller time scale compared to that of fluid dynamics. For long-pulse laser-induced cavitation, the same approach is no longer valid. It would not be able to predict the effects of the lasting energy input, such as the possible continuation of phase transition and the formation of non-spherical bubbles. In this work, we couple the multiphase compressible inviscid Navier-Stokes equations with a laser radiation equation that models the absorption of laser by the fluid flow. The laser radiation equation is obtained by customizing the radiative transfer equation (RTE) using the special properties of laser, including monochromaticity, directionality, and a measurable (often non-zero) focusing or diverging angle. The key components of the computational framework include an embedded boundary method that allows the solution of laser and fluid governing equations on the same mesh, a method of latent heat reservoir for vaporization prediction, a local level set method for interface tracking, and the FIVER (FInite Volume method with Exact multi-material Riemann solvers) method to enforce interface conditions. The algorithms and properties of this framework were recently published in the Journal of Computational Physics, together with some verification tests \citep{zhao2023simulating}. The FIVER method by itself is the pivot of a body of literature that includes numerical method development, verification and validation, and various applications in aerospace, ocean, and biomedical engineering (\citet{farhat2012fiver,main2017enhanced,ma2023efficient,islam2023fluid,huang2018family}, and the references therein). Compared to the physical model presented in \citet{zhao2023simulating}, a few improvements are made in this work, such as the inclusion of heat diffusion and the modeling of laser fiber using an embedded boundary method.

In two separate laboratory experiments, we use Holmium:Yttrium-Aluminum-Garnet (Ho:YAG) and Thulium fiber lasers to generate a pear-shaped bubble and an elongated bubble. In both cases, the bubble dynamics is recorded by high-speed optical imaging. In addition, the temporal profile of laser power is measured using a light detector. These experimental measurements are treated as ground truth in this study. We simulate the two experiments using the computational model described above. The measured laser power profile is used as an input to each simulation, which starts with a single phase (liquid water) in the entire computational domain. The simulations are capable of predicting bubble nucleation due to laser radiation. They provide transient, full-field results of laser radiance, temperature, pressure, velocity, and density. They also track the dynamics of the vapor bubble using a level set function. To validate the computational model, we compare the bubble dynamics predicted by the simulations with the high-speed images obtained from the experiments. Then, we analyze the full-field simulation results to explain the bubble dynamics and energy transmission. Based on the results, we hypothesize that the bubble's shape is determined by a race between advection and phase transition. At any time instant, if the speed of advection is higher than that of vaporization, the bubble tends to grow spherically. Otherwise, it tends to elongate along the laser beam direction. To clarify this hypothesis, we build a simplified model problem for which the aformentioned two speeds can be analytically defined. Then, we test the hypothesis using our simulation results.

The remainder of this paper in organized as follows. Section~\ref{sec:physical_model} describes the physical model adopted in this study, including governing equations, constitutive models, and a phase transition model. Section~\ref{sec:comp_framework} provides a summary of the numerical methods used to solve the model equations. In Sections~\ref{sec:validation_Ho} and~\ref{sec:validation_TFL}, we present the experimental and simulation results for a pear-shaped bubble and an elongated bubble. In Section~\ref{sec:transition}, we discuss the transition between these two different shapes. Finally, Section~\ref{sec:conclusion} provides a summary of this study and some concluding remarks. 

\section{Physical model}
\label{sec:physical_model}

\subsection{Fluid dynamics and thermodynamics} 
\label{sec:govering_equations}
\begin{figure}
  \centerline{\includegraphics[width = 0.98\textwidth]{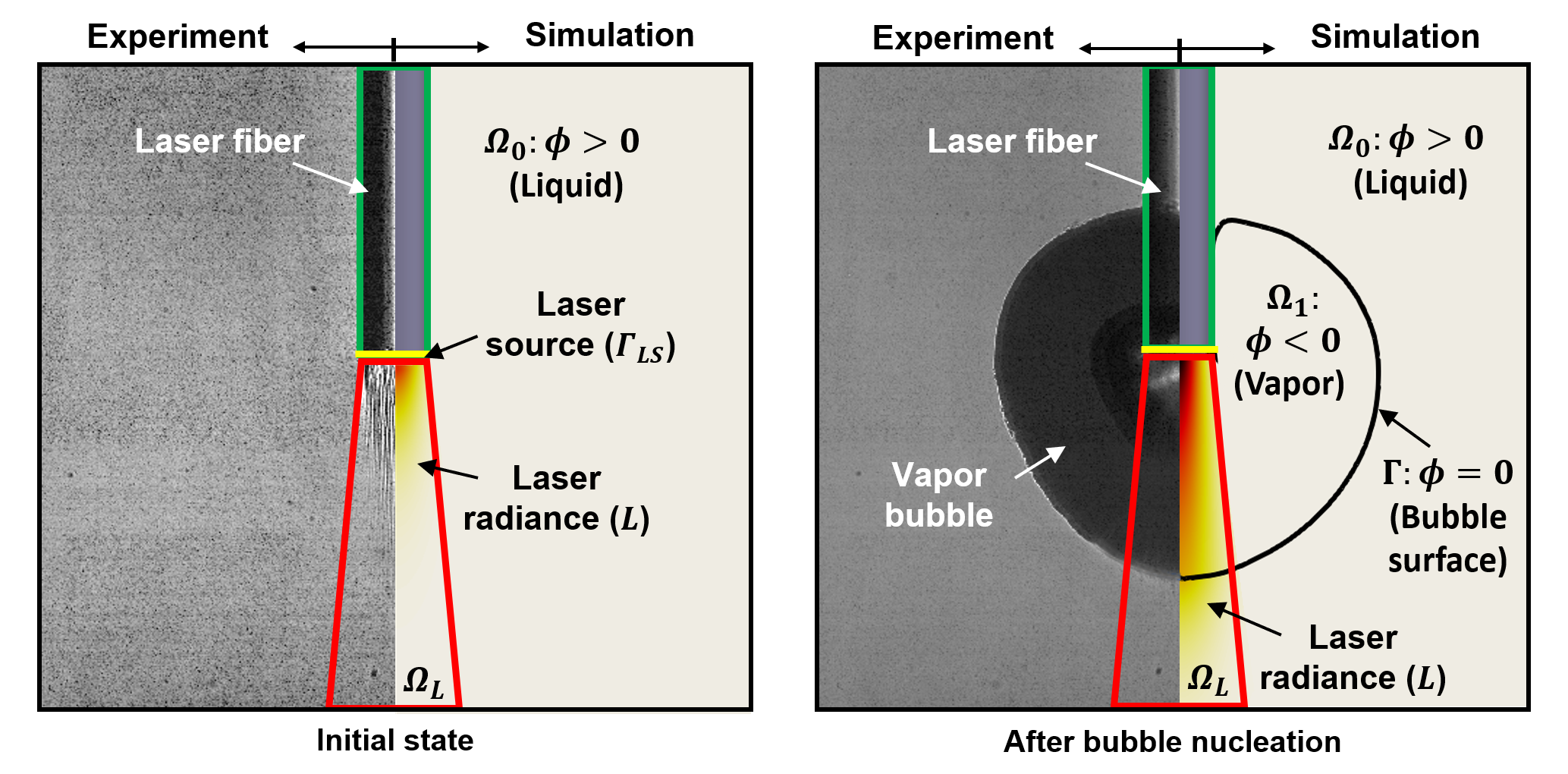}}
  \caption{Long-pulse laser-induced vaporization and bubble expansion: An example problem.}
 \label{fig:computational_model}
\end{figure}

Figure~\ref{fig:computational_model} illustrates the problem investigated in this paper, showing a test case that generates a pear-shaped vapor bubble. The computational analysis is designed to start at the time when laser is just activated. At this time, the fluid domain is completely occupied by liquid water. The analysis is expected to predict the localized water vaporization due to laser radiation, the subsequent bubble and fluid dynamics, and the dissipation of laser energy in this two-phase fluid medium. Therefore, we solve the following compressible inviscid Navier-Stokes equations that include radiative heat transfer.
\begin{equation}
\frac{\partial \bm{W}(\bm{x},t)}{\partial t} + \nabla \cdot \mathcal{F}(\bm{W}) = \nabla \cdot \mathcal{G}(\bm{W}), \qquad \forall \bm{x} \in \Omega = \Omega_0 \cup \Omega_1,~t>0,
\label{eq:NSEquation}
\end{equation}
with
\begin{equation}
\bm{W}=\begin{bmatrix} \rho \\ \rho \bm{V} \\ \rho e_{t} \end{bmatrix}, \qquad \mathcal{F} = \begin{bmatrix} \rho \bm{V}^{T} \\  \rho \bm{V} \otimes \bm{V} + p \bm{I} \\ (\rho e_{t} + p)\bm{V}^T \end{bmatrix},\qquad \mathcal{G} = \begin{bmatrix} \bm{0}^T \\ \bm{0}\\ (k\nabla T- \bm{q_r})^T  \end{bmatrix}.
\label{eq:NSEquation_terms}
\end{equation}

Here, $\Omega\subset\mathbb{R}^3$ denotes the domain of the fluid flow. Open subsets $\Omega_0$ and $\Omega_1$ represent the subdomains occupied by the liquid and vapor phases, respectively. They are time-dependent. In most experiments, a sharp boundary between the liquid and vapor phases can be clearly captured by high-speed cameras. Therefore, we assume $\Omega_0 \cap \Omega_1 = \emptyset$.~$\rho$, $\bm{V}$, $p$, and $T$ denote the fluid's density, velocity, pressure, and temperature, respectively. $e_t$ is the total energy per unit mass, given by
\begin{equation}
e_t = e + \dfrac{1}{2} |\bm{V}|^2,
\end{equation}
where $e$ denotes the fluid's internal energy per unit mass. $k$ is the thermal conductivity coefficient, which takes different values in $\Omega_0$ and $\Omega_1$. $\bm{q_r}$ denotes the radiative heat flux induced by laser. Equation~\eqref{eq:NSEquation} needs to be closed by a complete equation of state (EOS) for each phase, including a temperature equation. The computational model and solver utilized in this study supports arbitrary convex EOS \citep{ma2023efficient}. In this study, we adopt the Noble-Abel stiffened gas equation \citep{le2016noble} for both phases. Specifically,
\begin{equation}
p_{\mathcal{I}}(\rho,e) = \left( \gamma_\mathcal{I}  - 1 \right) \dfrac{e-q_\mathcal{I}}{\dfrac{1}{\rho}   - b_\mathcal{I}}  - \gamma_\mathcal{I} p_{c\mathcal{I}},
\label{eq:NASG_EOS}
\end{equation}
in which the subscript $\mathcal{I} \in \{0, 1\}$ identifies the liquid ($0$) and vapor ($1$) phases. For each phase, $\gamma$, $p_c$, $q$, and $b$ are constant parameters that characterize its thermodynamic properties. For example, a non-zero $b$ allows the model to have a variable Gr\"{u}neisen parameter that depends on $\rho$. Clearly, \eqref{eq:NASG_EOS} is a generalization of perfect gas, stiffened gas, and Noble-Abel equations of state \citep{le2016noble}. 

For a given pressure equation like~\eqref{eq:NASG_EOS}, the choice of temperature equation is not unique. We adopt the one proposed in \citet{le2016noble}, i.e.,
\begin{equation}
T_\mathcal{I}(\rho,e) = \dfrac{1}{c_{v\mathcal{I}}} \Big(e-q_\mathcal{I} -\big(\dfrac{1}{\rho}  -b_\mathcal{I}\big)p_{c\mathcal{I}}\Big),
\label{eq:Temperature_law}
\end{equation}
where $c_v$ denotes the specific heat capacity at constant volume, assumed to be a constant. It can be shown that the specific heat capacity at constant pressure, $c_p$, is also a constant, given by $c_p = \gamma c_v$. Combining \eqref{eq:NASG_EOS} and \eqref{eq:Temperature_law} gives a complete EOS that satisfies the first law of thermodynamics.

Two groups of EOS parameter values are tested in this study, as shown in Table~\ref{tab:EOS_params}. Neither of them was calibrated specifically for laser-induced cavitation \citep[see]{le2016noble,zein2013modeling}. We will show in Section~\ref{sec:validation_Ho} that the simulation result is indeed influenced by these parameter values.

\begin{table}
  \begin{center}
  \def~{\hphantom{0}}
  \begin{tabular}{cccccccc}
      Group & Reference  & Phase  &   $\gamma$ & $p_c\text{(Pa)}$ & $c_v\text{(J/(kg K))}$  & $b\text{(m}^3\text{/kg)}$ &$q\text{(J/kg)}$  \\[3pt]
      \hline
       \multirow{2}{*}{1} & Zein \etal   & Liquid & ~~2.057 & ~~1.066$\times 10^9$ & ~3.449$\times10^3$ & ~~0~~ & -1994.674$\times10^3$~\\
         & (2013)  & Vapor & ~~1.327 &  ~~0~~~  & ~1.2~$\times10^3$ & ~~0~~ & 1995$\times10^3$\\
       \hline
       \multirow{2}{*}{2} & Métayer \etal   & Liquid & ~~1.19~ & ~~7.028$\times 10^8$ & ~4.285$\times10^3$ & 6.61$\times10^{-4}$ & -1177.788$\times10^3$~\\
         &  (2016) & Vapor & ~~1.47~ &  ~~0~~~  & ~0.955$\times10^3$ & ~~0~~ & 2077.616$\times10^3$\\
  \end{tabular}
  \caption{Noble-Abel stiffened gas EOS parameters for water.}
  \label{tab:EOS_params}
  \end{center}
\end{table}

\subsection{Liquid-vapor interface} 

We model the bubble surface as a sharp interface with zero thickness. It is defined by
\begin{equation}
\Gamma = \partial \Omega_0 \cap \partial \Omega_1.
\end{equation}

On the interface, we assume continuity of normal velocity and pressure, i.e. 
\begin{equation}
\left.
\begin{array}{l}
\Big(\lim\limits_{\bm{x}'\rightarrow\bm{x} ,~ \bm{x}'\in\Omega_0} \bm{V}(\bm{x}',t) - \lim\limits_{\bm{x}'\rightarrow\bm{x} ,~  \bm{x}'\in \Omega_1} \bm{V}(\bm{x}',t) \Big)\cdot \bm{n}(\bm{x},t) = 0, \\ 
\lim\limits_{\bm{x}'\rightarrow \bm{x} ,~  \bm{x}'\in \Omega_0} p(\bm{x}',t) = \lim\limits_{\bm{x}'\rightarrow \bm{x} ,~  \bm{x}'\in \Omega_1} p(\bm{x}',t),
\end{array} 
\right.
\qquad \forall \bm{x}\in \Gamma,~t\geq 0,
\label{eq:liquid_gas_interface}
\end{equation}
where $\bm{n}$ denotes the normal to $\Gamma$.

 $\Gamma$ is time-dependent, and must be solved for during the analysis. We represent it implicitly as the zero level set of a signed distance function, $\phi$, defined in the closure of $\Omega$. That is,
\begin{equation}
\Gamma(t) = \{\bm{x}\in\overline{\Omega},~\phi(\bm{x},t)=0\},
\end{equation}
where $\overline{\Omega}$ denotes the closure of $\Omega$.

In this way, the aforementioned phase identifier, $\mathcal{I}$, is given by
\begin{equation}
    \mathcal{I}(\bm{x},t) = 
    \begin{cases}
    0, & \text{if } \phi(\bm{x},t)>0\},\\
    1, & \text{if } \phi(\bm{x},t)<0\}.
    \end{cases}
    \label{eq:identification_number}
\end{equation}

The evolution of $\Gamma$ in time is driven by both phase transition and advection. The advection of $\Gamma$ by the fluid flow is governed by the level-set equation,
\begin{equation}
\frac{\partial \phi}{\partial t} + \bm{V} \cdot \nabla \phi = 0,
\label{eq:levelset}
\end{equation}
where $\bm{V}$ is the flow velocity.

At the beginning of the analysis, $\Omega = \Omega_0$, and $\Gamma = \emptyset$. Therefore, we initialize $\phi$ to be a constant positive value everywhere in the domain, and start solving~\eqref{eq:levelset} only after phase transition starts. The detection and handling of laser-induced phase transition will be discussed in Section~\ref{sec:phase_transition}.

\subsection{Laser radiation}
\label{sec:laser_radiation}

Let~$\Omega_L\subset\Omega$ denote the region in the fluid domain that is exposed to laser. The laser generators used in this study have a flat surface with a small diverging angle. Therefore, $\Omega_L$ is in the shape of a truncated cone  (Fig.~\ref{fig:computational_model}). Within $\Omega_L$, energy conservation implies
\begin{equation}
\nabla \cdot \big(\mathcal{L}\hat{\bm{s}}\big) = \mu_{\alpha}(\eta) \mathcal{L}_{b}( \bm{x}, \eta) - \mu_{\alpha}(\eta)\mathcal{L}(\bm{x}, \hat{\bm{s}}, \eta) - \mu_{s}(\eta) \mathcal{L}(\bm{x}, \hat{\bm{s}}, \eta) + \dfrac{\mu_{s}(\eta)}{4 \pi} \int_{4 \pi} \mathcal{L}(\bm{x}, \bm{\hat{s}_i}, \eta) \Phi(\bm{\hat{s}_i},\bm{\hat{s}}) d \bm{\hat{s}}_i,
\label{eq:RTE}
\end{equation}
where $\mathcal{L}=\mathcal{L}(\bm{x},\hat{\bm{s}}, \eta)$ denotes the spectral radiance (dimension: [mass][time]\textsuperscript{-3}[length]\textsuperscript{-1}) at position $\bm{x}\in\mathbb{R}^3$, along direction $\hat{\bm{s}}$, at wavelength $\eta$. $\mu_{\alpha}$ and $\mu_{s}$ denote the coefficients of absorption and scattering, respectively. They depend on the laser's wavelength and the medium. $\mathcal{L}_{b}$ denotes the blackbody radiance. $\Phi(\bm{\hat{s}_i},\bm{\hat{s}})$ is the scattering phase function, which gives the probability that a ray from one direction $\bm{\hat{s}_i}$ would be scattered into another direction $\bm{\hat{s}}$. If $\hat{\bm{s}}$ is independent of $\bm{x}$, the left-hand side of Eq.~\eqref{eq:RTE} simplifies to $\nabla \mathcal{L} \cdot \bm{s}$, which gives the well-known radiative transfer equation (RTE) \citep[][]{modest2013radiative,howell2020thermal}.

The radiative heat flux $\bm{q}_r$ in Eq.~\eqref{eq:NSEquation} is obtained by integrating $\mathcal{L}$ over all directions and the interval of relevant wavelengths, $(\eta_{\text{min}},~\eta_{\text{max}})$. That is,
\begin{equation}
\bm{q_{r}} (\bm{x}) = \int_{\eta_{\text{min}}}^{\eta_{\text{max}}} \int_{4\pi} \mathcal{L}(\bm{x}, \hat{\bm{s}}, \eta)  \bm{\hat{s}} ~d\hat{\bm{s}} d\eta.
\label{eq:radiative_flux_def}
\end{equation}

The special properties of laser light allows us to simplify~\eqref{eq:RTE} and~\eqref{eq:radiative_flux_def}. The intensity of the laser light is much higher than the blackbody radiance. Therefore, we assume
\begin{equation}
\mathcal{L}_b(\bm{x},\eta)=0.
\label{eq:blackbody_radiance}
\end{equation}

Also, $\mathcal{L}$ is non-zero only along the direction of laser propagation ($\bm{s}$) and at the fixed laser wavelength. That is, 
\begin{equation}
\mathcal{L}(\bm{x},\hat{\bm{s}},\eta) = L(\bm{x})\delta(\hat{\bm{s}}-\bm{s})\delta(\eta-\eta_0),
\label{eq:laser_radiance}
\end{equation}
where $\delta(\cdot)$ denotes the delta function, and the variable $L(\bm{x})$ on the right-hand side is radiance (dimension: [mass][time]\textsuperscript{-3}). With these assumptions, a laser radiation equation is obtained by substituting~Eq.~\eqref{eq:laser_radiance} and~Eq.~\eqref{eq:blackbody_radiance}  into Eq.~\eqref{eq:RTE}, i.e.,
\begin{equation}
\nabla\cdot(L\bm{s}) = \nabla L \cdot\bm{s} + (\nabla\cdot\bm{s}) L = - \mu_\alpha L.
\label{eq:laser_eq}
\end{equation}

Here, $\bm{s} = \bm{s}(\bm{x})$ is a known function, but not a constant unless the laser light has a parallel beam. Similarly, substituting Eq.~\eqref{eq:laser_radiance} into Eq.~\eqref{eq:radiative_flux_def} yields 
\begin{equation}
\bm{q}_r = L \bm{s}.
\label{eq:laser_radiativeflux}
\end{equation}

In this study, we measure the time history of laser power, $P_0(t)$, in laboratory experiments. We assume the laser radiance on the surface of the laser fiber (i.e.,~$\Gamma_{LS}$ in Fig.~\ref{fig:computational_model}) follows a Gaussian distribution, also known as a Gaussian beam~\citep{welch2011optical}. Therefore, on $\Gamma_{LS}$ we have
\begin{equation}
L(\bm{x},t) = \frac{2P_0(t)}{\pi w_0^2} \exp\Big( \frac{-2 \big|\bm{x}-\bm{x}_0\big|^2}{w_0^2} \Big),
\label{eq:Gaussian_beam_eq}
\end{equation}
where $\bm{x}_0$ denotes the center of $\Gamma_{LS}$, and $w_0$ is the waist radius.~\eqref{eq:Gaussian_beam_eq} serves as a Dirichlet boundary condition for the laser radiation equation,~\eqref{eq:laser_eq}.

\subsection{Phase transition}
\label{sec:phase_transition}

\begin{figure}
  \centerline{\includegraphics[width = 0.98\textwidth]{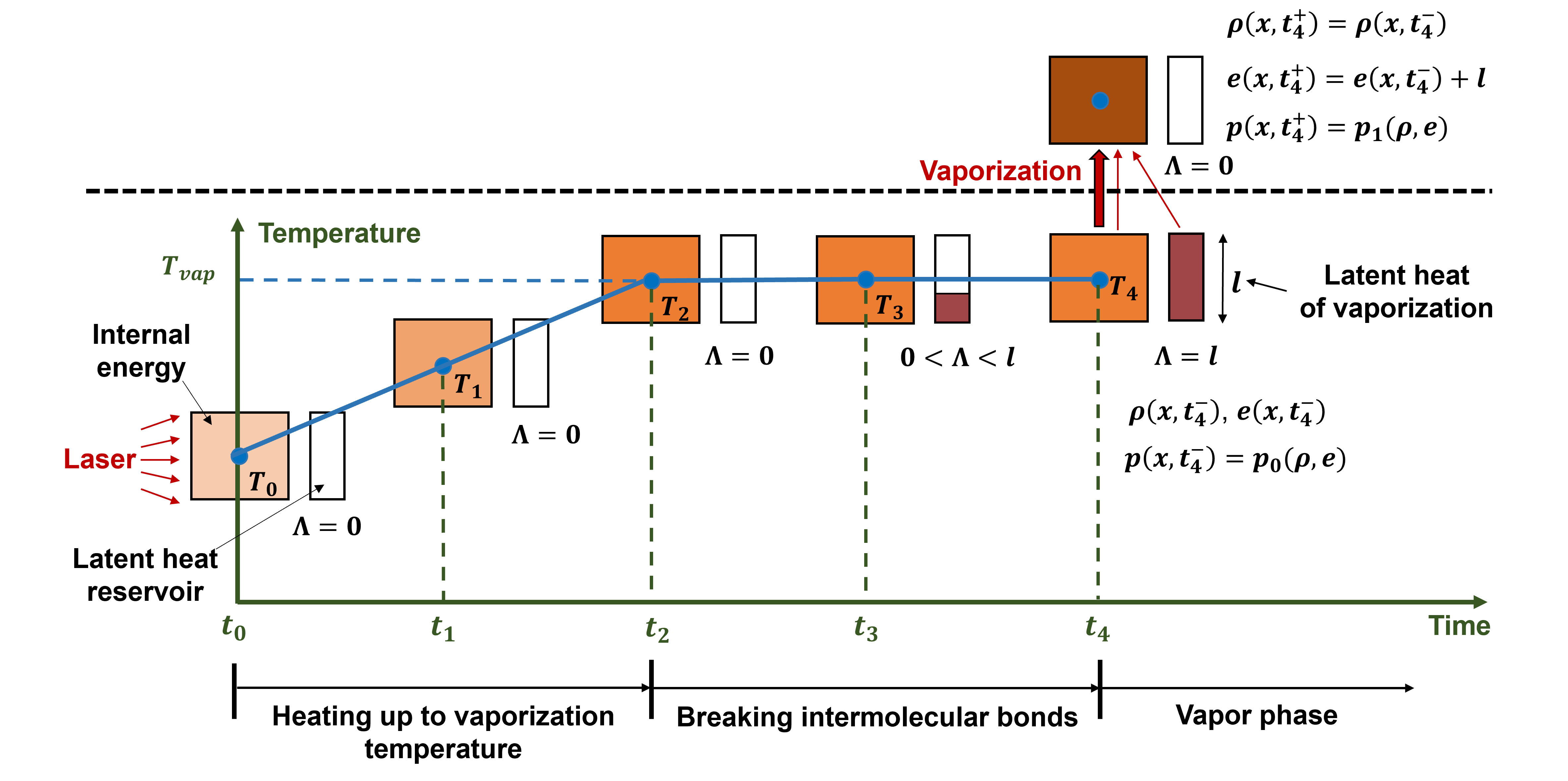}}
  \caption{Predicting laser-induced vaporization by the method of latent heat reservoir.}
 \label{fig:phase_transition_model}
\end{figure}

We employ the method of latent heat reservoir proposed in~\citet{zhao2023simulating} to predict laser-induced vaporization. As shown in Fig.~\ref{fig:phase_transition_model}, the fundamental idea of this method is to introduce a new variable, $\Lambda(\bm{x},t)$, to track the intermolecular potential energy in the liquid phase. The vaporization temperature, $T_{\text{vap}}$, and latent heat, $l$, are assumed to be constant and used in the method as coefficients. When the analysis starts, $\Lambda(\bm{x},t)$ is initialized to $0$ everywhere. As the liquid absorbs laser energy, temperature increases gradually. At any point $\bm{x}\in \Omega_0$, once $T_{\text{vap}}$ is reached, additional heat --- due to radiation, advection, or diffusion --- is added to $\Lambda$. When $\Lambda$ reaches $l$, phase transition occurs. In Fig.~\ref{fig:phase_transition_model}, this time instant is denoted by $t_4$. We assume that  phase transition completes instantaneously at each point through an isochoric process. The state variables after phase transition are given by

\begin{align}
\mathcal{I}^+ &= 1,\label{eq:I_correction}\\
\rho^+ &= \rho^-,\\
e^+ &= e^- + \Lambda^-,\\
\Lambda^+ &= 0,\\
\phi^+ &= -\dfrac{\Delta x}{2},\label{eq:phi_correction}
\end{align}
where the superscripts $-$ and $+$ indicate the variable's value before and after phase transition, e.g.,
\begin{equation}
\rho^{+} \equiv \lim\limits_{t\rightarrow t_4 \atop t>t_4}\rho(\bm{x},t).
\end{equation}

In \eqref{eq:phi_correction}, $\Delta x$ denotes the local element size in the computational mesh. The  pressure and temperature after phase transition are obtained naturally from the EOS of the vapor phase, i.e.,
\begin{align}
p^+ &= p_1(\rho^+, e^+),\\
T^+ &= T_1(\rho^+, e^+).
\label{eq:phase_transition}
\end{align}

The pressure rise, $p^+ - p^-$, drives the expansion of the vapor bubble.

At each time step, if phase transition occurs, the level set function $\phi$ is restored to a signed distance function by solving the reinitialization equation,
\begin{equation}
\dfrac{\partial\phi}{\partial \tilde{t}} + S(\phi_0)(|\nabla\phi| - 1) = 0,
\label{eq:reinit}
\end{equation}
to the steady state. Here, $\tilde{t}$ is a fictitious time variable. $\phi_0$ is the level set function before reinitialization. $S(\phi_0)$ is a smoothed sign function, given by
\begin{equation}
S(\phi_0) = \dfrac{\phi_0}{\sqrt{\phi_0^2 + \varepsilon^2}},
\end{equation}
where $\varepsilon$ is a constant coefficient, set to the minimum element size of the mesh. The steady state solution of Eq.~\eqref{eq:reinit} is then used as the new initial condition to integrate the level set equation~\eqref{eq:levelset} forward in time.

\section{Numerical methods}
\label{sec:comp_framework}

\subsection{FIVER (FInite Volume method based on Exact multiphase Riemann problem solvers)}
\label{sec:FIVER}
\begin{figure}
  \centerline{\includegraphics[width = 0.7\textwidth]{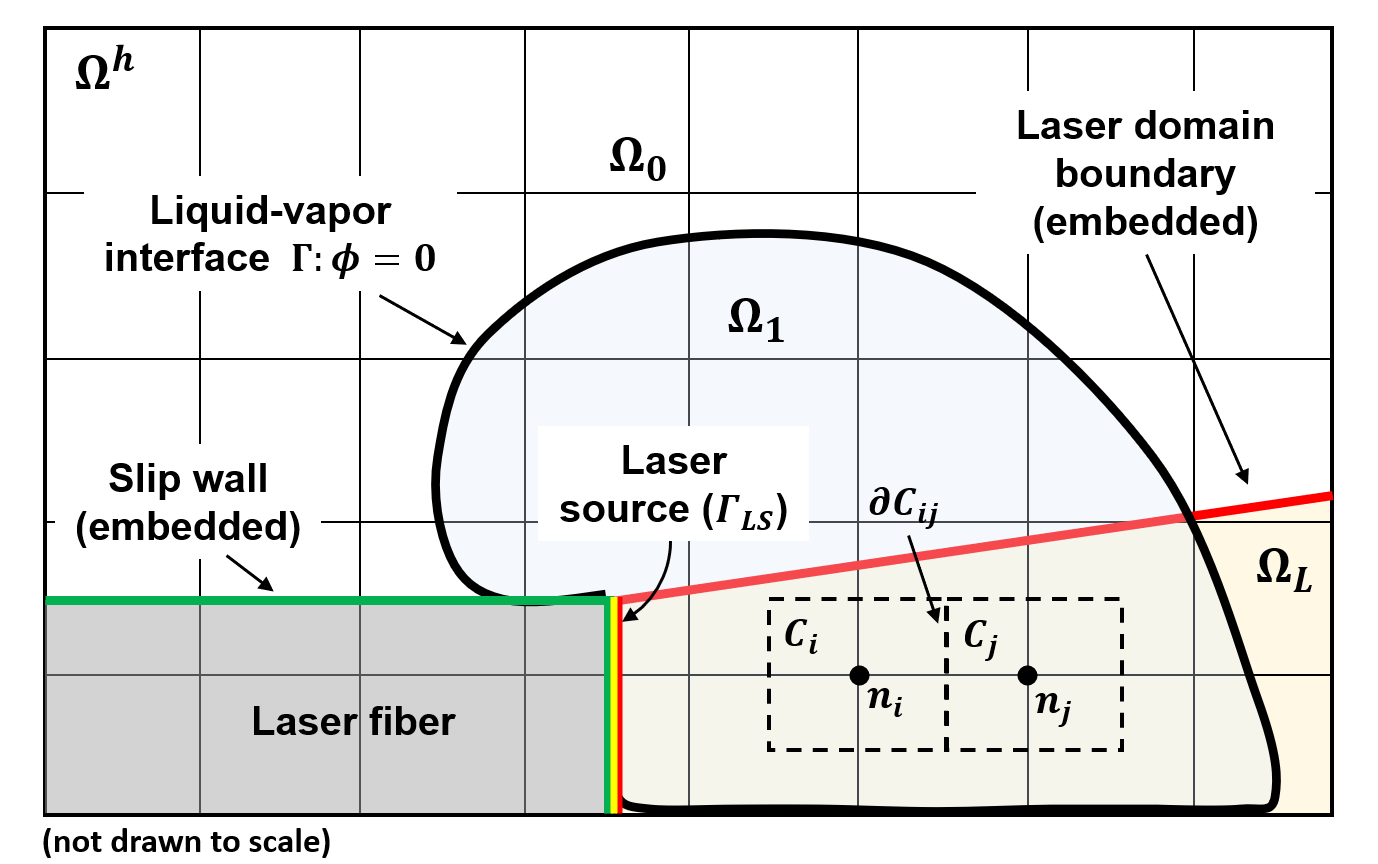}}
  \caption{Finite volume discretization of the spatial domain.}
 \label{fig:discretization}
\end{figure}

We apply a recently developed laser-fluid computational framework to solve the above model equations \citep{zhao2023simulating}. The fluid governing equations are semi-discretized using a fixed, non-interface conforming finite volume mesh, denoted by $\Omega^h$ (Fig.~\ref{fig:discretization}). The laser fiber is modeled as a fixed slip wall, and the associated boundary condition is enforced using an embedded boundary method \citep{cao2018robin,cao2021spatially,wang2017multiphase,cao2021shock}. Therefore, $\Omega^h$ also covers the region occupied by the laser fiber. Around each node $\bm{n}_i\in\Omega^h$, a control volume $C_i$ is created. Applying the standard finite volume spatial discretization to Eq.~\eqref{eq:NSEquation} yields
\begin{equation}
\frac{\partial \bm{W}_i}{\partial t} + \frac{1}{|\, {C_i} \,|} \sum_{j \in Nei(i)} \int_{\partial C_{ij}} \mathcal{F}(\bm{W}) \cdot \bm{n}_{ij} dS = \frac{1}{|\, {C_i} \,|} \int_{C_i} \nabla\cdot \mathcal{G}(\bm{W}) d\bm{x},
\label{eq:NS_discretized}
\end{equation} 
where $\bm{W}_i$ denotes the average value of $\bm{W}$ in $C_i$. $Nei(i)$ denotes the set of nodes connected to node $\bm{n}_i$. $\partial C_{ij} = \partial C_i \cap \partial C_j$. $\bm{n}_{ij}$ is the unit normal to $\partial C_{ij}$. $|C_i|$ denotes the volume of $C_i$.

The FIVER method is used to calculate the advective flux across each facet $\partial C_{ij}$. Depending on the locations of nodes $\bm{n}_i$ and $\bm{n}_j$, there are four different cases.
\begin{enumerate}[(a)]
\item If $\bm{n}_i$ and $\bm{n}_j$ belong to the same phase subdomain ($\Omega_0$ or $\Omega_1$), the conventional monotonic upstream-centered scheme for conservation laws (MUSCL) is used to compute the flux across $\partial C_{ij}$.
\item If $\bm{n}_i$ and $\bm{n}_j$ belong to different phase subdomains, a one-dimensional bimaterial Riemann problem is constructed along the edge between $\bm{n}_i$ and $\bm{n}_j$, i.e.,
\begin{equation}
\frac{\partial \bm{w}}{\partial \tau} + \frac{\partial \mathcal{F}(\bm{w})}{\partial \xi} = 0, \quad \text{with} \quad \bm{w}(\xi,0) = 
\begin{cases}
&\bm{w}_i, \qquad \text{if} \ \xi \leq 0, \\
&\bm{w}_j, \qquad \text{if} \ \xi > 0, 
\end{cases}
\label{eq:1DRiemannProblem}
\end{equation}
where $\tau$ denotes the time coordinate. $\xi$ denotes the spatial coordinate along the axis aligned with $\bm{n}_{ij}$ and centered at the phase interface between $\bm{n}_i$ and $\bm{n}_j$. The initial state $\bm{w}_i$ and $\bm{w}_j$ are the projections of $\bm{W}_i$ and $\bm{W}_j$ on the $\xi$ axis. This 1D two-phase compressible flow problem can be solved exactly \citep{ma2023efficient}. The solution is self-similar, and satisfies the interface conditions~\eqref{eq:liquid_gas_interface}. The states to the left and right of the interface are used in the numerical flux function to compute the advective fluxes across $\partial C_{ij}$.
\item If one of $\bm{n}_i$ and $\bm{n}_j$ belongs to a fluid phase subdomain ($\Omega_0$ or $\Omega_1$), while the other one is covered by the laser fiber, a 1D half-Riemann problem is constructed and solved exactly \citep{cao2021shock,wang2017multiphase}. The solution at the wall boundary is used to compute the advective fluxes across $\partial C_{ij}$.
\item If both $\bm{n}_i$ and $\bm{n}_j$ are covered by the laser fiber, the flux across $\partial C_{ij}$ is set to zero.
\end{enumerate}

\subsection{Diffusive heat fluxes}

The right-hand side of Eq.~\eqref{eq:NS_discretized} includes two parts, namely the heat diffusion term ($\displaystyle\int_{C_i} \nabla\cdot (k \nabla T) d\bm{x}$) and the radiation term ($\displaystyle\int_{C_i} \nabla\cdot \bm{q_r} d\bm{x}$). The heat diffusion term is evaluated using a finite volume method, i.e.
\begin{equation}
    \int_{C_i} \nabla\cdot (k \nabla T) d\bm{x} = \int_{\partial C_{ij}} (k \nabla T) \cdot \bm{n}_{ij} dS = \sum\limits_{j\in Nei(i)} (k \nabla T)_{ij} A_{ij} \cdot \bm{n}_{ij},
    \label{eq:heat_diffusion_discretization}
\end{equation}
where $A_{ij}$ is the area of facet $\partial C_{ij}$. $(k \nabla T)_{ij}\cdot \bm{n}_{ij}$ denotes the heat flow due to diffusion cross the facet $\partial C_{ij}$. In this work, it is computed by 
\begin{equation}
    (k \nabla T)_{ij}\cdot \bm{n}_{ij} = k_{ij} \frac{T_i - T_j}{\|\bm{x}_i - \bm{x}_j\|_2},
    \label{eq:heat_diffusion_scheme}
\end{equation}
with
\begin{equation}
k_{ij} = \dfrac{2k_i k_j}{k_i + k_j}.
\label{eq:interface_conductivity}
\end{equation}

Here, $k_i$ and $k_j$ denote the thermal conductivity at nodes $\bm{n}_i$ and $\bm{n}_j$, specifically. We assume that the liquid-vapor interface is isothermal, and it can be shown that if $\bm{n}_i$ and $\bm{n}_j$ belong to different phase subdomains, Eqs.~\eqref{eq:heat_diffusion_scheme} and~\eqref{eq:interface_conductivity} enforce energy balance at the interface, with interface temperature \citep{faghri2006transport}
\begin{equation}
    T_{ij} = \frac{k_i T_i + k_j T_j}{k_i + k_j}.
\end{equation}

We set the thermal conductivity at all the nodes covered by the laser fiber to zero. In this way, Eqs.~\eqref{eq:heat_diffusion_scheme} and~\eqref{eq:interface_conductivity} enforce the adiabatic boundary condition at the surface of the laser fiber.

\subsection{laser radiation and laser-fluid coupling}

The radiation term is evaluated by 
\begin{equation}
\int_{C_i} (\nabla\cdot \bm{q_r}) d\bm{x} = (\nabla\cdot \bm{q_r})_i {|\, {C_i} \,|}.
\label{eq:radiation_discretized}
\end{equation}

At each time step, the divergence of the radiative flux, $\nabla \cdot \bm{q_r}$, is computed using the current solution of the laser radiation equation~\eqref{eq:laser_eq}. In this work, the laser radiation equation is discretized using the same finite volume mesh ($\Omega^h$) created for the fluid governing equations (Fig.~\ref{fig:discretization}). Specifically, integrating \eqref{eq:laser_eq} within an arbitrary cell $C_i$ yields
\begin{equation}
\sum\limits_{j\in Nei(i)} A_{ij}L_{ij} (\bm{s}_{ij}\cdot \bm{n}_{ij})  = - {|\, {C_i} \,|} \mu_{\alpha}(T_i)L_i,
\label{eq:FVM3D}
\end{equation}
where $L_{i}$ is the cell-average of $L$ within $C_i$. $\bm{s}_{ij}$ denotes the laser direction at $\partial C_{ij}$, which is set to the laser direction at the midpoint between nodes $\bm{n}_i$ and $\bm{n}_j$.  $L_{ij}$ is the value of $L$ at facet $\partial C_{ij}$. In this work, it is evaluated by the mean flux method, i.e.
\begin{equation}
L_{ij} = 
\begin{cases}
\alpha L_i + (1-\alpha) L_j & \text{if } \bm{s}_{ij} \cdot n_{ij} \geq 0,\\
(1-\alpha) L_i + \alpha L_j & \text{if } \bm{s}_{ij} \cdot n_{ij} < 0,\\
\end{cases}
\label{eq:mean_flux_method}
\end{equation} 
where $\alpha\in [0.5, 1]$ is a numerical parameter. Substituting Eq.~\eqref{eq:mean_flux_method} into Eq.~\eqref{eq:FVM3D} yields a system of linear equations with the cell-averages of laser radiance as unknowns. The Gauss-Seidel method is applied to solve this system to get $L_i$. Then, $\nabla \cdot \bm{q_r}$ is obtained by
\begin{equation}
 (\nabla \cdot \bm{q_r})_i  =  \nabla \cdot (L_i \bm{s}) = - \mu_{\alpha}(T_i) L_i.
\label{eq:NS_source_term}
\end{equation}

When solving the laser radiation equation, the fluid mesh $\Omega^h$ does not contain a subset of nodes, edges, and elements that resolve the boundary of $\Omega_L$ or the laser propagation directions $\bm{s}(\bm{x})$. In this work, the embedded boundary method proposed in \cite{zhao2023simulating} is adopted to resolve this issue. This method involves the population of ghost nodes outside the side boundary of the laser domain using second-order mirroring and interpolation techniques.

\vspace{1cm}
The numerical methods described above are implemented in the M2C solver \citep{Wang2021M2C}, which is used to run the simulations reported in this paper. Several verification studies can be found in \citet{islam2023fluid,zhao2023simulating,cao2021shock}, and earlier papers cited therein. An outline of the solution procedure within each time step is provided below. For simplicity, the Forward Euler method is assumed here for time integration. In the actual simulations presented in this paper, a second-order Runge-Kutta method is used.\\

\begin{tabularx}{\textwidth}{cll}
\bf{Input:} & \multicolumn{2}{l} {Numerical solution at $t^n$: $\bm{W}^n$, $\mathcal{I}^n$, $\phi^n$, $L^n$, and $\Lambda^n$.}\\
(1) & \multicolumn{2}{l} {Compute the residual of the Navier-Stokes equations (Eq.~\eqref{eq:NS_discretized}).}\\
 & (1.1) & Compute the advective fluxes (Sec.~\ref{sec:FIVER}).\\
 & (1.2) & Compute the diffusive heat fluxes (Eq.~\eqref{eq:heat_diffusion_discretization}).\\
 & (1.3) & Compute the radiative heat source (Eq.~\eqref{eq:radiation_discretized}).\\
(2) & \multicolumn{2}{l} {Advance the fluid state by one time step $\Rightarrow$ $\bm{W}^{n+1}$, $\Lambda^{n+1}$. Apply boundary }\\
& \multicolumn{2}{l} {conditions.}\\
(3) & \multicolumn{2}{l} {Compute the residual of the level set equation (Eq.~\eqref{eq:levelset}).}\\
(4) & \multicolumn{2}{l} {Advance the level set function by one time step $\Rightarrow$ $\phi^{n+1}$. Apply boundary }\\
& \multicolumn{2}{l} {conditions.}\\
(5) & \multicolumn{2}{l} {Update phase identifier using $\phi^{n+1}$ $\Rightarrow$ $\mathcal{I}^{n+1}$. Update fluid state ($\bm{W}^{n+1}$, $\Lambda^{n+1}$) at}\\
 & \multicolumn{2}{l}{ nodes that changed phase due to interface motion.}\\
(6) & \multicolumn{2}{l} {Check for phase transition (Sec.~\ref{sec:phase_transition}). Update $W^{n+1}$, $\Lambda^{n+1}$, $\mathcal{I}^{n+1}$, and $\phi^{n+1}$ at }\\
&  \multicolumn{2}{l} {nodes that have undergone phase transition (Eqs.~\eqref{eq:I_correction}-\eqref{eq:phase_transition}).}\\
(7) & \multicolumn{2}{l} {If phase transition occurred, reinitialize the level set equation (Eq.~\eqref{eq:reinit}).}\\
(8) & \multicolumn{2}{l} {Solve the laser radiance equation for $L^{n+1}$ (Eq.~\eqref{eq:FVM3D}).}\\
\bf{Output:} & \multicolumn{2}{l} {Numerical solution at $t^{n+1}$: $\bm{W}^{n+1}$, $\mathcal{I}^{n+1}$, $\phi^{n+1}$, $L^{n+1}$, and $\Lambda^{n+1}$.}
\end{tabularx}

%

\section{A pear-shaped bubble}
\label{sec:validation_Ho}

We employ the computational framework described above to simulate a laboratory experiment that generates a pear-shaped bubble. The key parameters of the simulation, including laser fiber diameter, laser absorption coefficient, and the divergence angle of the laser beam, are set to match the setup of the experiment. The laser power used in the simulation (i.e. $P_0(t)$ in Eq.~\eqref{eq:Gaussian_beam_eq} is determined by fitting the laser power profile measured in the experiment. The simulation's output includes the time-histories of the density, temperature, pressure, velocity, and laser radiance fields of the two-phase flow, and the level-set function that tracks the liquid-vapor interface. The bubble's nucleation and evolution are compared with high-speed optical images obtained from the experiment. By analyzing the full-field results obtained from the simulation, we try to investigate the causal relation between the laser setting and the bubble's shape. 

\subsection{Comparison of experimental and numerical results}

\subsubsection{Laboratory experiment}
\label{sec:HoYAG_experiment}

\begin{figure}
  \centerline{\includegraphics[width = 0.5\textwidth]{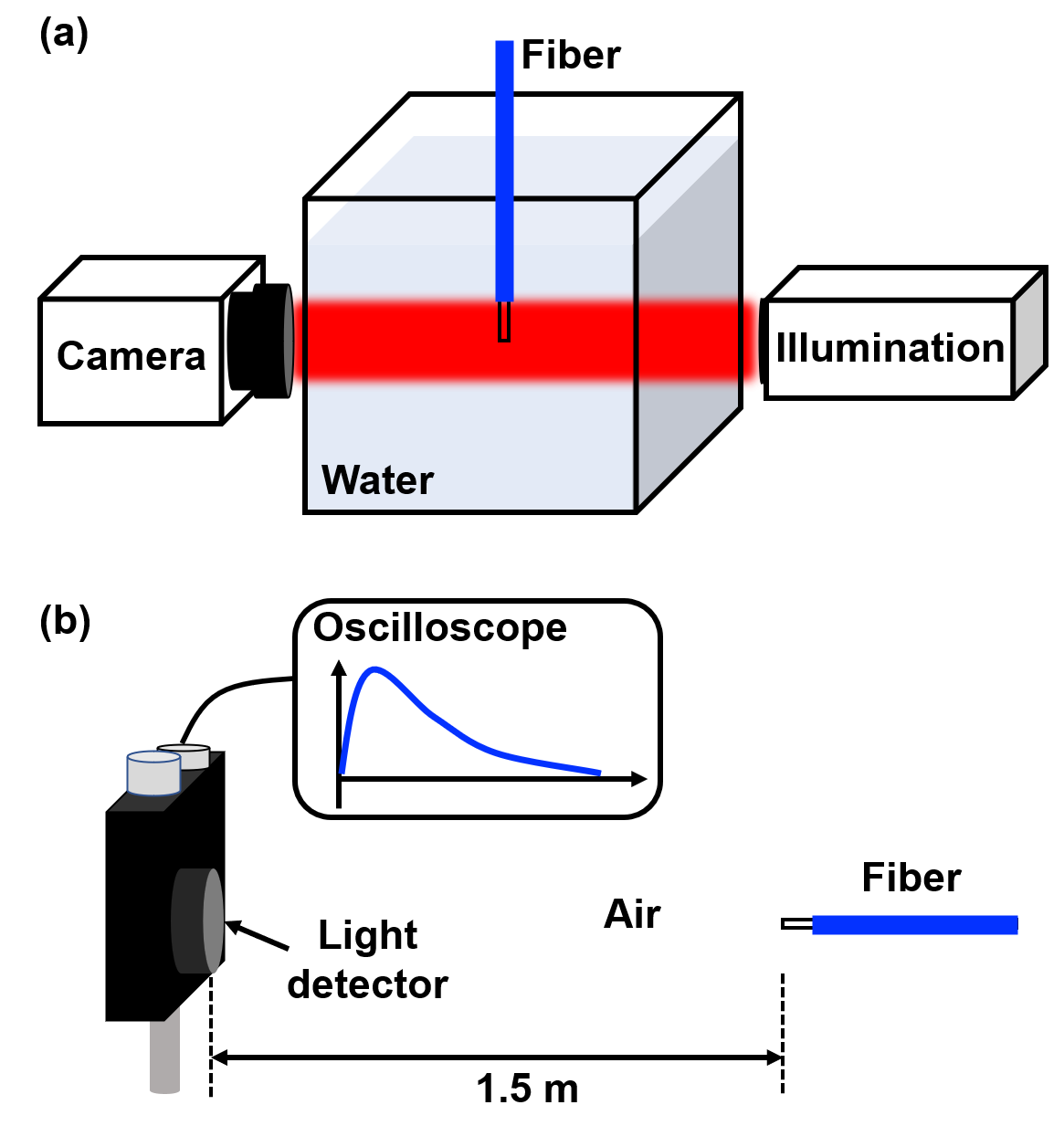}}
  \caption{Schematic representation of the experimental setup for laser-induced cavitation.
  (\textit{a}) Setup for capturing the bubble dynamics, and (\textit{b}) Setup for measuring the temporal profile of laser power.}
 \label{fig:experiment_setup_Ho}
\end{figure}

In this experiment, a commercial Ho:YAG laser lithotripter (H Solvo $35$-watt laser, Dornier MedTech, Munich, Germany) with a wavelength of $2080~\text{nm}$ is used to generate the vapor bubble. It is operated at the energy level of $0.2~\text{J}$ with a pulse duration of $70~\text{\textmu s}$, measured at full width at half maximum. It is clearly a long-pulsed laser, as the acoustic time scale is less than $1~\text{\textmu s}$. Figure~\ref{fig:experiment_setup_Ho}(a) shows a schematic representation of the experimental setup. To facilitate the delivery of the pulsed laser, an optical fiber (Dornier SingleFlex $400$, numerical aperture: $0.26$, Munich, Germany) with a core diameter of $0.365~\text{mm}$ is used. The fiber directs laser into a transparent acrylic container ($150 \times 150 \times 300 ~\text{mm}^3$) filled with degassed water. During the experiment, a series of high-speed images are captured using an ultrahigh-speed camera. To enable direct shadowgraph imaging, a $10$-ns pulsed laser system (SI-LUX-$640$, Specialised Imaging) provides the required illumination.

To measure the laser power profile, an additional procedure is conducted in air. The laser pulse is directed towards a light detector (PDA$10$D, Thorlabs, Newton, NJ) positioned at a distance of $1.5$ m, as illustrated in Fig.~\ref{fig:experiment_setup_Ho}(b). The light detector converts the received light into an electronic signal, which is displayed on an oscilloscope. Since air has minimal absorption of laser energy, the recorded data can be considered a reliable indication of the laser power output from the laser fiber when generating the vapor bubble in the bulk fluid.

\begin{figure}
  \centerline{\includegraphics[width = 0.95\textwidth]{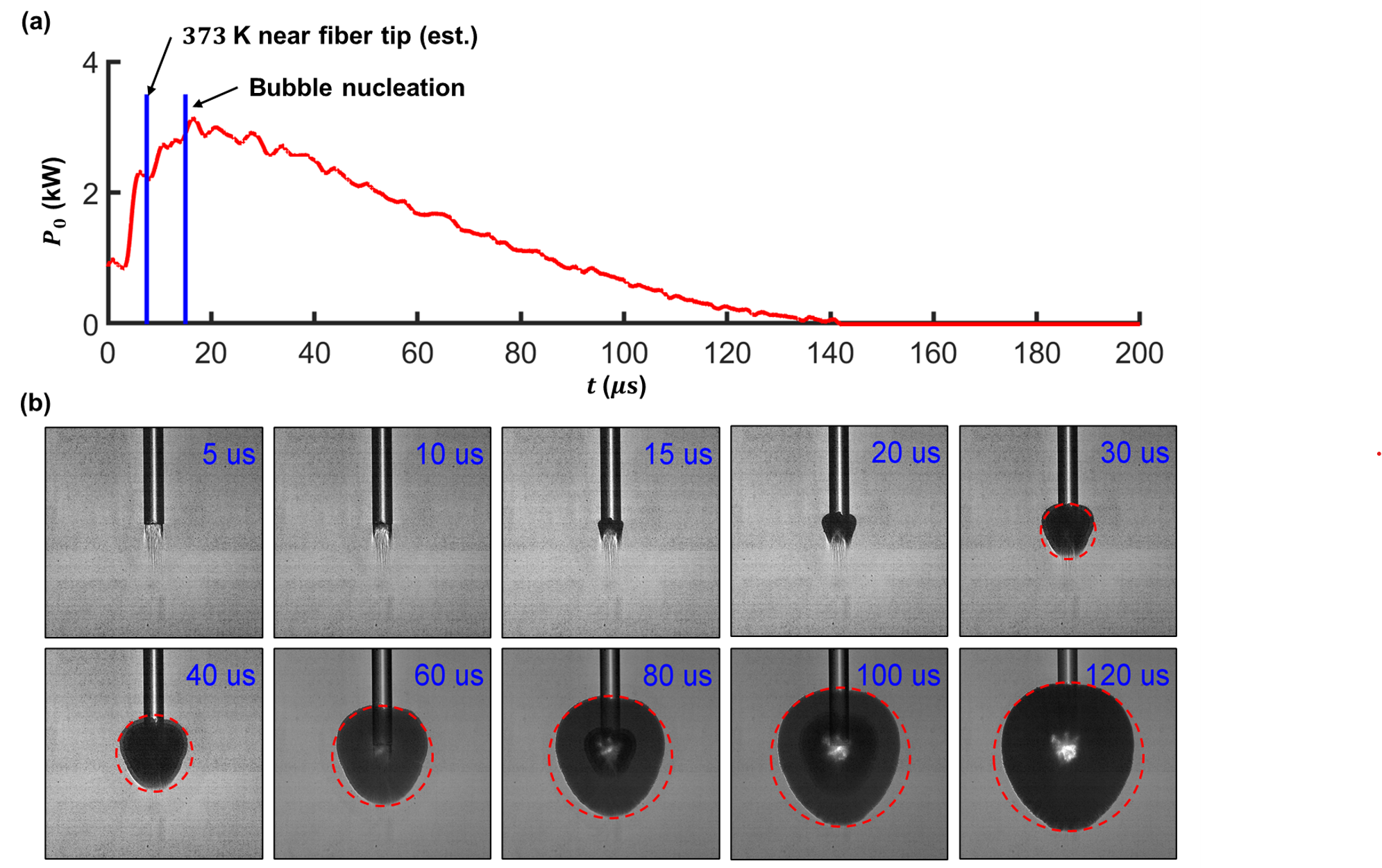}}
  \caption{Experimental results obtained with a Ho:YAG laser.
  (\textit{a}) Laser pulse profile measured in air, and (\textit{b}) dynamics of the vapor bubble in the bulk fluid. Perfect circles are drawn in sub-figure (\textit{b}) to show that the vapor bubble is not spherical.}
 \label{fig:experiment_result_Ho}
\end{figure}

Figure~\ref{fig:experiment_result_Ho}(a) presents the time-history of the laser power measurement. The laser power increases from the beginning and reaches the maximum at approximately $17~\text{\textmu s}$. After that, it gradually drops to zero. The graph reveals some fluctuations, which are attributed to measurement noise. Integrating the measured laser power in time yields a total pulse energy of $0.2$ J, which is in good agreement with the specified energy level. Figure~\ref{fig:experiment_result_Ho}(b) presents a series of high-speed images that show the nucleation and evolution of a vapor bubble at the tip of the laser fiber. The bubble becomes visible at $15~\text{\textmu s}$. It expands continuously, eventually acquiring a pear-like shape. The bubble maintains cylindrical symmetry about the central axis of the laser beam. However, it is not spherical, unlike the bubbles obtained from previous experiments that use short-pulsed lasers (e.g.,~\cite{brujan_nahen_schmidt_vogel_2001, zhong2020model,lauterborn2013shock}).

Using the measured laser power profile, we can estimate the time it takes to heat water near the laser fiber tip to the vaporization temperature, $T_{\text{vap}}$. Here, we assume $T_{\text{vap}} = 373.15~\text{K}$. We define a cylindrical region next to the laser fiber tip, with a diameter of $0.365~\text{mm}$ (the laser fiber diameter) and a small depth of $0.1~\text{mm}$. The energy required to raise water temperature in this region from the room temperature (assumed to be $293.15~\text{K}$) to $T_{\text{vap}}$ can be estimated by
\begin{equation}
    \Delta E = c_v \Delta T \rho V \approx 0.003 ~\text{J},
\end{equation}
with $c_v = 4.285\times 10^3~\text{J/(kg~K)}$, and $\Delta T = 80~\text{K}$. The percentage of laser energy absorbed in this region can be roughly estimated by the Beer-Lambert law, which is the one-dimensional version of~\eqref{eq:laser_eq}. That is,
\begin{equation}
    \frac{\partial{L}}{\partial x} = -\mu_{\alpha} L.
\end{equation}

The solution of this equation is simply $\displaystyle L(x) = L_0 e^{-\mu_\alpha x}$, where $L_0$ denotes the laser radiance at the source (i.e., $x=0$). Setting $\mu_{\alpha} = 2.42~\text{mm}^{-1}$ for the Ho:YAG laser~\citep{fried2018advances}, we find that approximately $21.5\%$ of the laser energy is absorbed within the depth of the cylindrical region defined above. Now, by integrating the measured laser power profile (Fig.~\ref{fig:experiment_result_Ho}(a)), we can estimate that at approximately $9~\text{\textmu s}$, the temperature within the small cylindrical region reaches $T_{\text{vap}}$. From the high-speed images, the vapor bubble does not appear until $15~\text{\textmu s}$. This finding implies a clear delay in bubble nucleation, which can be attributed to the high latent heat of water.

\subsubsection{Numerical simulation}
\label{sec:HoYAG_simulation}

\begin{figure}
  \centerline{\includegraphics[width = 0.95\textwidth]{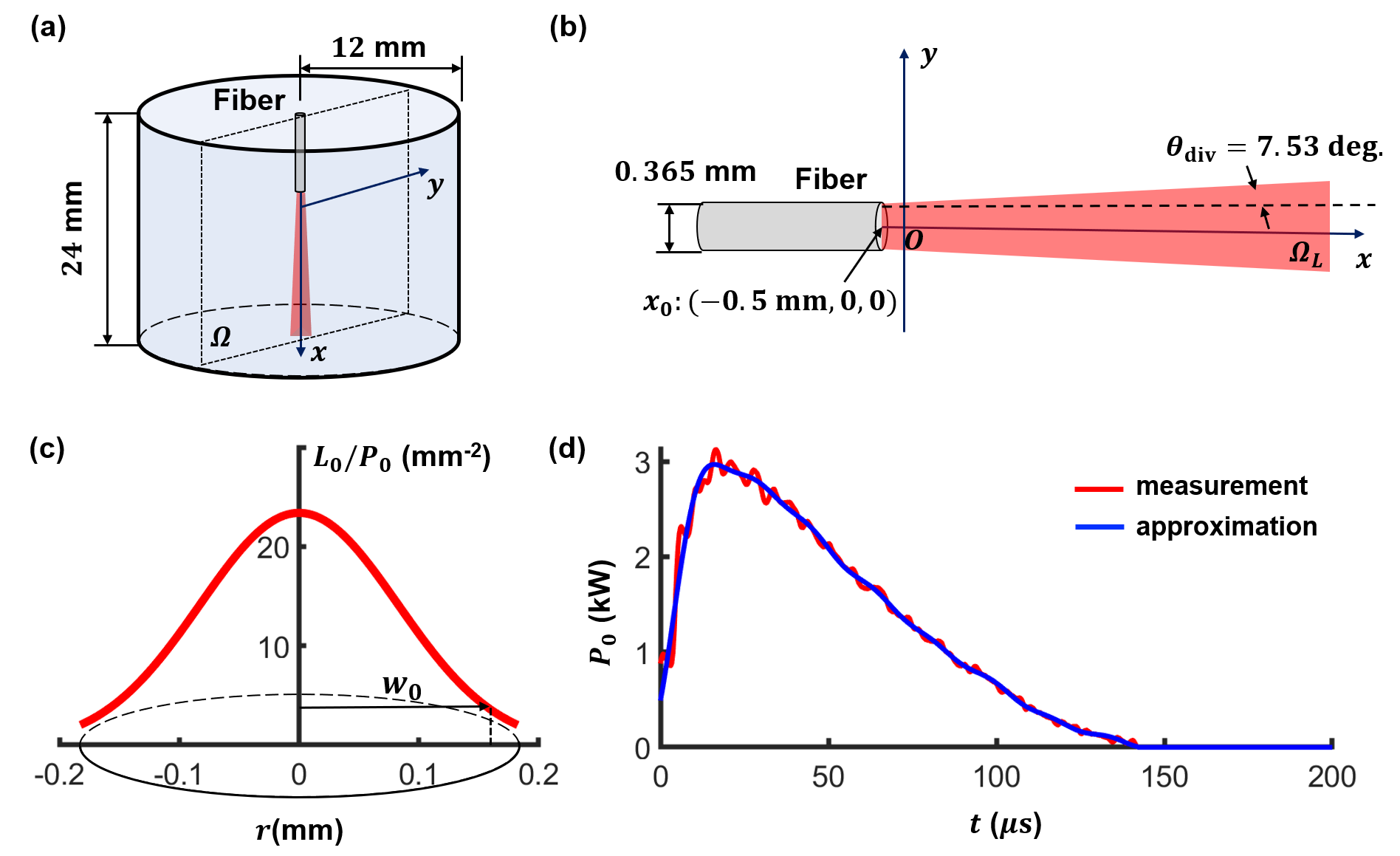}}
  \caption{Vapor bubble generated by a Ho:YAG laser: Simulation setup.
   (\textit{a}) Spatial domain with cylindrical symmetry. (\textit{b}) Geometry of the laser radiation domain. (\textit{c}) Spatial profile of laser radiance on the laser source plane. (\textit{d}) Temporal profile of laser power.}
 \label{fig:numerical_setup_HoYAG}
\end{figure}

Figure~\ref{fig:numerical_setup_HoYAG} shows the setup of the simulation. A cylindrical domain with a radius of $12$ mm and a length of $24$ mm is adopted. It is initially occupied by liquid water with density $\rho_0 = 0.001~\text{g}/\text{mm}^3$, velocity $v_0=0~\text{mm}/\text{s}$, pressure $p_0 = 100~\text{kPa}$, and temperature $T_0 = 293.15 K$.

The laser source is positioned at $x=-0.5~\text{mm}$, as shown in Fig.~\ref{fig:numerical_setup_HoYAG}(b). The laser fiber is modeled as a rigid structure, embedded in the computational domain. It has a radius $r = 0.1825~\text{mm}$, which is consistent with the laboratory experiment. The laser beam has a divergence angle $\theta_{\text{div}} = 7.53^{\circ}$, which depends on the numerical aperture (NA) of the fiber and the laser wavelength \citep{pal2010guided}. The spatial profile of laser radiance on the source plane is specified as a Gaussian function (Eq.~\eqref{eq:Gaussian_beam_eq}) with waist radius $w_0 = 0.165~\text{mm}$, as shown in Fig.~\ref{fig:numerical_setup_HoYAG}(c). The temporal profile of laser power ($P_0$) is specified as the blue line shown in Fig.~\ref{fig:numerical_setup_HoYAG}(d). It is obtained by fitting the experimental measurement using fast Fourier transform (FFT). The pulse shape is approximately a triangle, with the power growing from $0$ to $2.98~\text{kW}$ within $24~\text{\textmu s}$, then vanishing gradually within $136~\text{\textmu s}$. The laser absorption coefficient, $\mu_\alpha$, is set to $2.42~\text{mm}^{-1}$ for liquid water \citep{fried2018advances} and $0.001~\text{mm}^{-1}$ for the vapor. The vaporization temperature and latent heat of vaporization are set by $T_{\text{vap}} = 373.15~\text{K}$ and $l = 2256.4~\text{J}/\text{g}$, respectively. 

The Nobel-Abel stiffened gas EOS (Eq.~\eqref{eq:NASG_EOS}) is employed to model both liquid water and water vapor. First, we adopt the EOS parameters presented in~\cite{zein2013modeling}, which are listed as Group $1$ in Table~\ref{tab:EOS_params}. The thermal conductivity, $k$, is set to $5.576\times 10^{-4}~\text{W/(mm~K)}$ for liquid water and $2.457\times 10^{-5}~\text{W/(mm~K)}$ \citep{wagner2010d2} for the vapor.

By the assumption of cylindrical symmetry, we solve the fluid governing equations on a two-dimensional mesh, while adding source terms to the equations to enforce the symmetry (see, e.g., \cite{islam2023fluid}). The mesh has approximately $2.4$ million rectangular elements. In the most refined area, the characteristic element size is about $1.5\times 10^{-3}~\text{mm}$. To put this into context, the diameter of the laser fiber is resolved by about $240$ elements. The local level set method with a bandwidth of $6$ elements on each side of the interface is used to track the vapor bubble. Both the fluid governing equations and the level set equation are integrated in time using a second-order Runge-Kutta method. The time step size is approximately $4\times 10^{-4}~\text{\textmu s}$. The simulation is terminated at $t = 120~\text{\textmu s}$, which roughly covers the bubble nucleation and expansion stage observed in the laboratory experiment. 

\begin{figure}
  \centerline{\includegraphics[width = 0.98\textwidth]{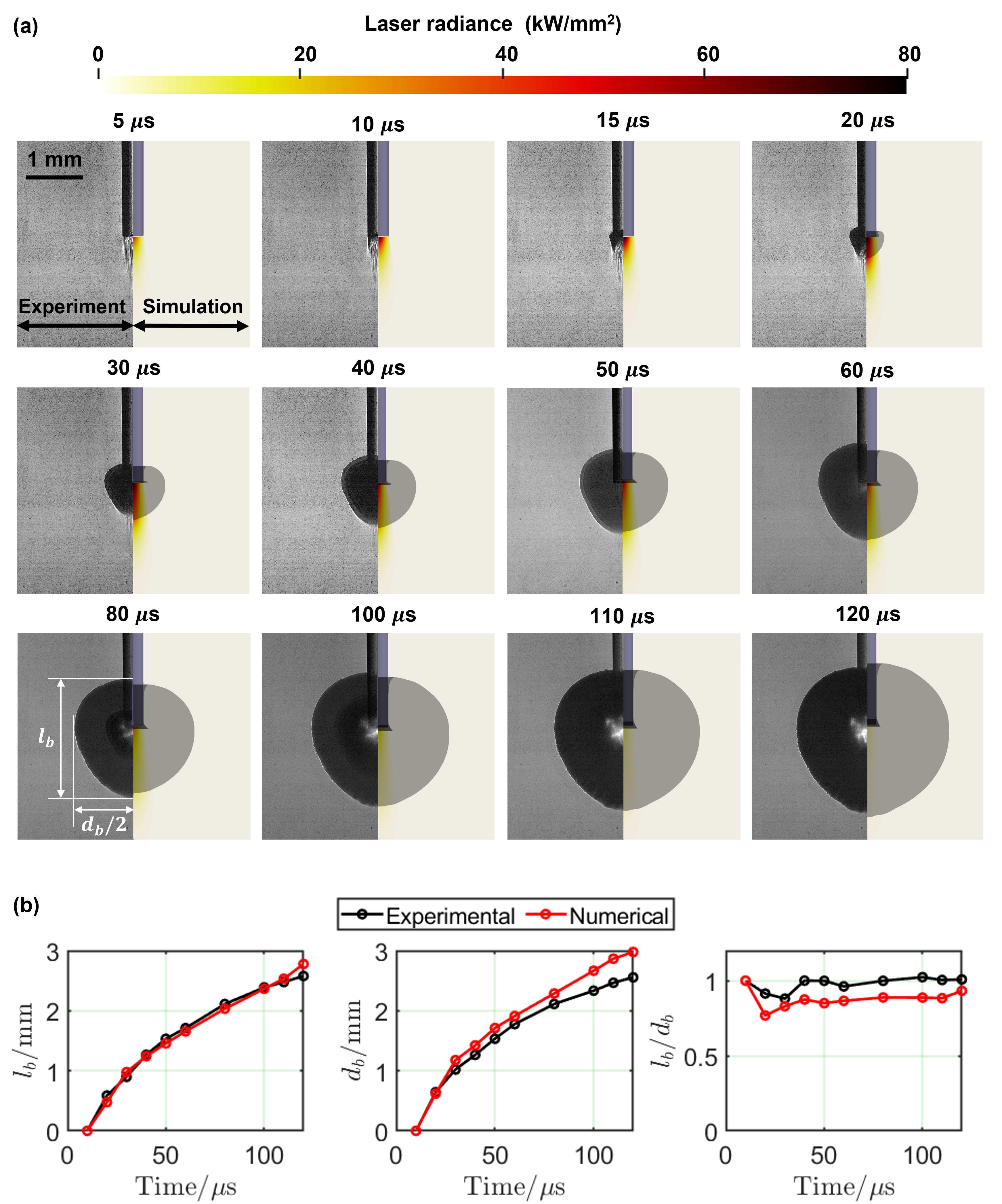}}
  \caption{Vapor bubble generated by a Ho:YAG laser: Comparison of bubble dynamics obtained from numerical simulation and laboratory experiment. (\textit{a}) Bubble nucleation and evolution. In each sub-figure, the left half shows the imaging result from the experiment, and the right half shows the bubble and laser radiance field predicted by the simulation. (\textit{b}) Evolution of bubble size and shape. $l_b$ and $d_b$ denote the maximum length of the bubble along and perpendicular to the laser fiber direction, respectively.}
 \label{fig:comp_HoYAG}
\end{figure}

The simulation also generates a pear-shaped bubble, as observed in the experiment. Figure~\ref{fig:comp_HoYAG} presents a detailed comparison between the experimental data and the simulation result. In sub-figure (a), the high-speed images obtained from the experiment and the simulation results (bubble dynamics and laser radiance) are shown side by side. The simulation predicts that the vapor bubble nucleates at approximately $17.4~\text{\textmu s}$. This is similar to the finding from the experiment, in which the bubble appears at $15~\text{\textmu s}$. The overall bubble dynamics predicted by the simulation --- that is, the evolution of the bubble's size and overall shape --- agrees reasonably well with the experimental data.

To make a quantitative comparison between the simulation result and the experimental data, we measure the maximum length of the bubble in two directions, that is, along and perpendicular to the laser fiber direction. These two measurements are denoted as $l_b$ and $d_b$, respectively, and plotted in Fig.~\ref{fig:comp_HoYAG}(b). The value of $l_b$ obtained from the simulation matches its experiment counterpart very well. The discrepancy between the simulation and the experiment in $d_b$ is a bit larger, between $4\%$ and $15\%$ at different time instants. Also, the bubble's expansion speed in the perpendicular direction is slightly larger in the simulation than in the experiment. Furthermore, the aspect ratio of the bubble, defined as $l_b/d_b$, is also calculated, and plotted in Fig.~\ref{fig:comp_HoYAG}(b). In both the experiment and the simulation, it varies between $0.75$ and $1$.

\subsection{Delay of bubble nucleation due to latent heat}
\label{sec:timedelay_Ho}

As mentioned in Sec.~\ref{sec:HoYAG_experiment}, the laboratory experiment reveals a delay of bubble nucleation, that is, the time of bubble nucleation is clearly after the time when the vaporization temperature ($T_{\text{vap}}$) is reached at the fiber tip. The same phenomenon is observed in the simulation.

\begin{figure}
  \centerline{\includegraphics[width = 0.98\textwidth]{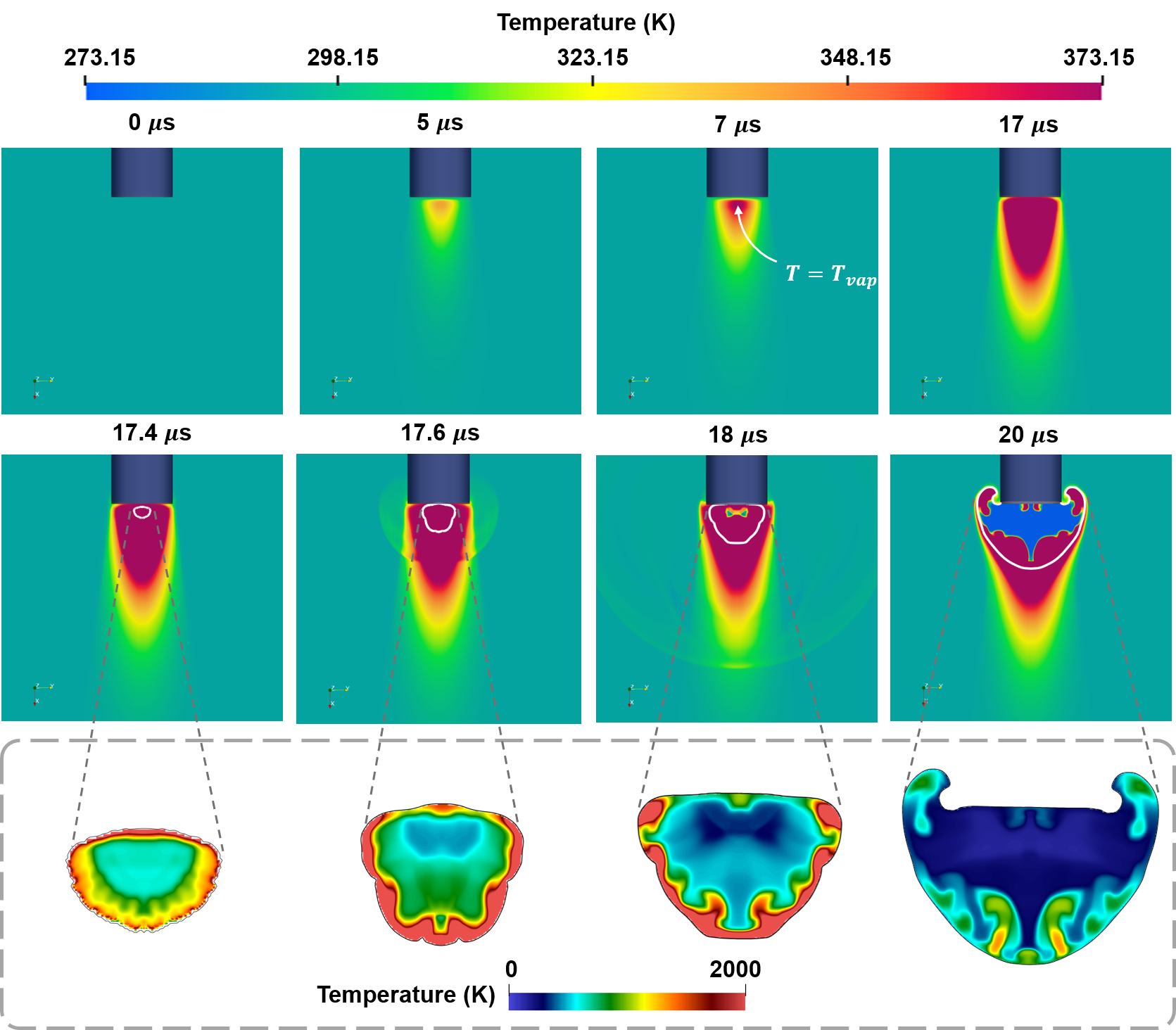}}
  \caption{Vapor bubble generated by a Ho:YAG laser: Temperature evolution in the first $20~\text{\textmu s}$. For the solutions between $17.4~\text{\textmu s}$ and $20~\text{\textmu s}$, a different color scheme and range is applied to show temperature variation inside the bubble.}
 \label{fig:temperature_HoYAG}
\end{figure}

Figure~\ref{fig:temperature_HoYAG} shows the temperature field at $8$ time instants during the early period of the simulation. At the beginning, the temperature of water in front of the laser fiber increases continuously, as it absorbs laser energy. This can be seen in the solution snapshots taken at $0$ to $7~\text{\textmu s}$. At $7~\text{\textmu s}$, the temperature of water next to the center of the fiber tip first reaches $T_{\text{vap}}$. This is about $10~\text{\textmu s}$ before bubble nucleation, which occurs at $17.4~\text{\textmu s}$. Within this time period, the region that reaches $T_{\text{vap}}$ expands continuously, but phase transition does not occur. 

\begin{figure}
  \centerline{\includegraphics[width = 0.8\textwidth]{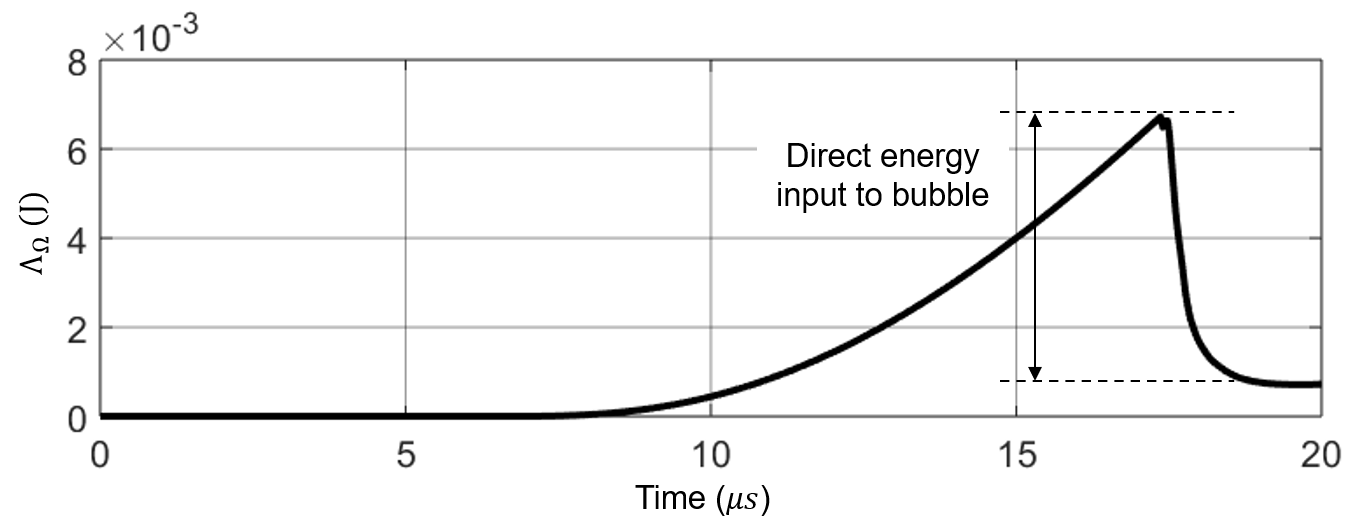}}
  \caption{Vapor bubble generated by a Ho:YAG laser: Time-history of the stored latent heat.}
 \label{fig:latentheat_HoYAG}
\end{figure}

This time delay is due to the fact that water has a high latent heat of vaporization. Based on the values of $l$ and $c_v$ adopted in the simulation, the latent heat is about $8$ times the energy needed to raise water temperature from $T_0$ to $T_{\text{vap}}$. Using the phase transition model described in Sec.~\ref{sec:phase_transition}, as soon as $T_{\text{vap}}$ is reached, any additional heat contributes towards increasing the intermolecular potential energy, thereby overcoming the required latent heat of vaporization. To examine this process more closely, we define
\begin{equation}
\Lambda_{\Omega} = \int_\Omega \rho\Lambda d\bm{x},
\end{equation}
which measures the total amount of latent heat in the computational domain. Figure~\ref{fig:latentheat_HoYAG} shows the time-history of $\Lambda_{\Omega}$.  At $6.9~\text{\textmu s}$, $\Lambda_{\Omega}$ becomes non-zero and begins to increase. Up to the time of bubble nucleation ($17.4~\text{\textmu s}$), the total energy stored in the latent heat reservoir is around $6.3$ mJ. By integrating the power profile (Fig.~\ref{fig:numerical_setup_HoYAG}(d)), we find that this value is approximately $16.84\%$ of the laser energy input up to the same time.

In this simulation, vaporization starts at $17.4~\text{\textmu s}$, and continues for a short period of time (less than $1~\text{\textmu s}$). Figure~\ref{fig:latentheat_HoYAG} shows that after vaporization stops, $\Lambda_{\Omega}$ drops to around $0.8$ mJ. Given that the total laser pulse energy is $0.2$ J, this result implies that only a small fraction of the laser energy input ($(6.3~\text{mJ} - 0.8~\text{mJ})/0.2~\text{J}=2.75\%$) is directly used to create the bubble.

After vaporization, the temperature inside the vapor bubble is up to $2000$ K, as shown in the solution snapshots at $17.4~\text{\textmu s}$ and $17.6~\text{\textmu s}$ in Fig.~\ref{fig:temperature_HoYAG}. The temperature near the bubble interface is higher than that in other regions, which is related to the direct energy input to the bubble after vaporization stops as depicted in Fig.~\ref{fig:latentheat_HoYAG}. Subsequently, the temperature inside the bubble gradually decreases as the bubble expands, owing to the combined effects of advection and diffusion.

\subsection{Generation of a pear-shaped bubble}
\label{sec:HoYAG_bubble}

\begin{figure}
  \centerline{\includegraphics[width = 0.98\textwidth]{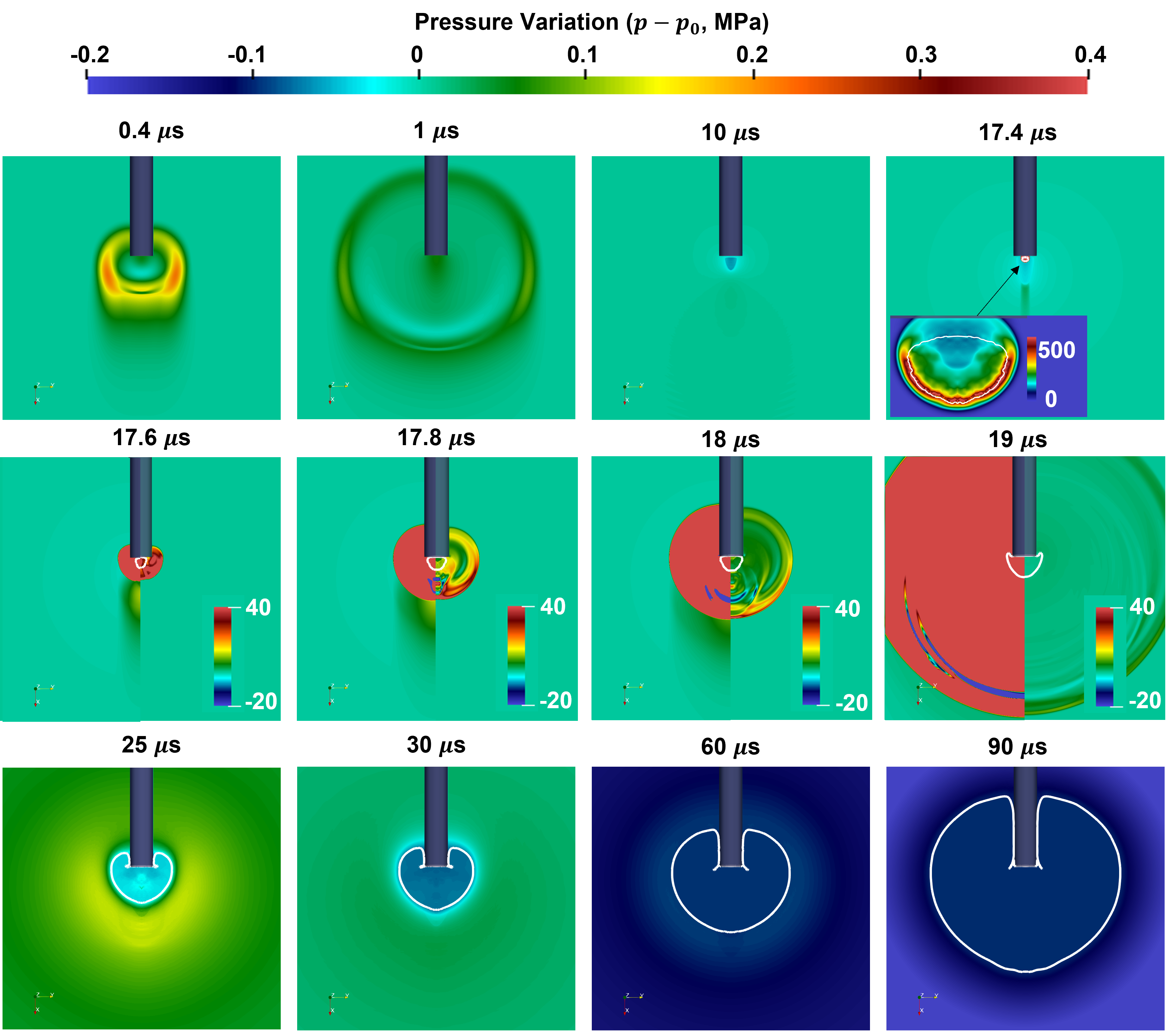}}
  \caption{Vapor bubble generated by a Ho:YAG laser: Evolution of the pressure field. For the solutions between $17.6~\text{\textmu s}$ and $19~\text{\textmu s}$, a different pressure range ($-20$ MPa to $40$ MPa) is applied to clearly show the pressure wave induced by the bubble.}
 \label{fig:pressure_HoYAG}
\end{figure}

\begin{figure}
  \centerline{\includegraphics[width = 0.98\textwidth]{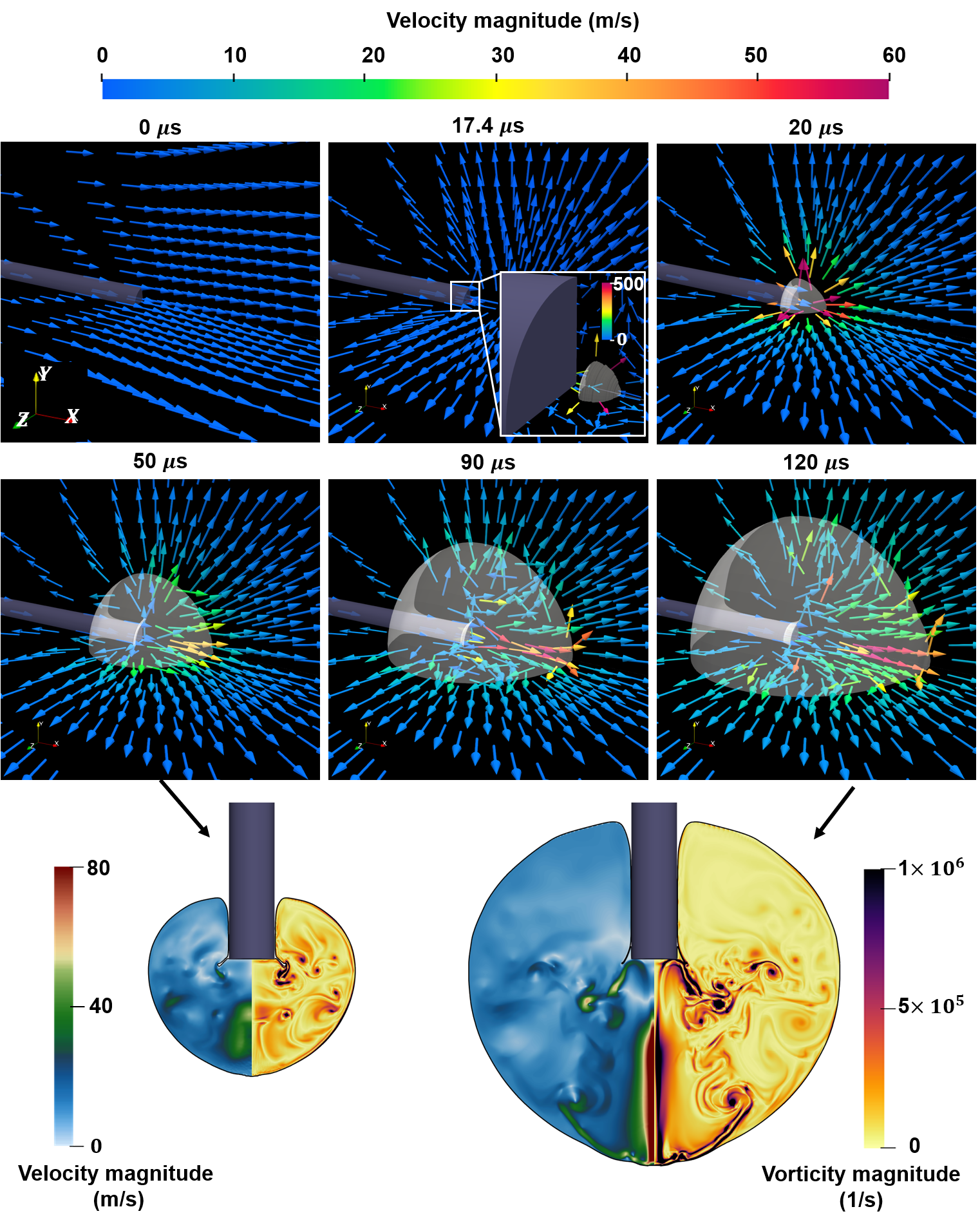}}
  \caption{Vapor bubble generated by a Ho:YAG laser: Evolution of the velocity field. The solution fields of velocity and vorticity magnitude inside the vapor bubble are shown for the time instances, $50~\text{\textmu s}$ and $120~\text{\textmu s}$, respectively.}
 \label{fig:velocity_HoYAG}
\end{figure}

In both the experiment and the simulation, a pear-shaped vapor bubble is obtained. To explain the formation of this shape, we look at the pressure and velocity fields obtained from the simulation  (Figures~\ref{fig:pressure_HoYAG} and~\ref{fig:velocity_HoYAG}). 

First, the two pressure snapshots at $0.4~\text{\textmu s}$ and $1~\text{\textmu s}$ (Fig.~\ref{fig:pressure_HoYAG}) capture an outgoing acoustic wave that emanated from the fiber tip. This is a thermal shock caused by the rapid increase of laser power. The vapor bubble nucleates at $17.4~\text{\textmu s}$. The pressure and velocity fields at this time are shown in Figs.~\ref{fig:pressure_HoYAG} and \ref{fig:velocity_HoYAG}. At this time, the bubble is already non-spherical. The pressure inside the bubble is high and nonuniform. Specifically, the pressure at the forward end is around $500$ MPa. The pressure at the backward end --- that is, near the fiber tip --- is much lower, around $100$ MPa. This pressure variation is a result of the brief continuation of vaporization. Regions that have just undergone phase transition have high pressures. Then, the pressure drops as the bubble expands. Therefore, the continuation of vaporization along the beam direction leads to the higher pressure at the forward end of the bubble. In summary, the simulation result at the time of bubble nucleation ($17.4~\text{\textmu s}$) already implies that the bubble will likely grow into a non-spherical shape, longer in the laser beam direction.

The high pressure inside the initial bubble also generates a weak shock wave that propagates outward at approximately the speed of sound in water. This wave can be seen in the pressure snapshots taken at $17.6~\text{\textmu s} - 19~\text{\textmu s}$ (Fig.~\ref{fig:pressure_HoYAG}). Afterwards, the pressure field becomes quiet, and the bubble continues growing due to inertia. The velocity field at $50~\text{\textmu s} - 120~\text{\textmu s}$ (Fig.~\ref{fig:velocity_HoYAG}) shows that the velocity around the bubble is nonuniform. It is higher at the forward end of the bubble compared to other regions. Furthermore, the velocity distribution within the bubble exhibits significant non-uniformity, as evidenced by the snapshots captured at time instances of $50~\text{\textmu s}$ and $120~\text{\textmu s}$. The velocity is notably higher in the vicinity of the central axis of the fiber, particularly near the forward end. This also explains the formation of a pear-like shape, instead of a sphere.

\subsection{Effect of the choice of EOS}
\label{sec:HoYAG_EOS}

We show that the result obtained from our laser-fluid coupled computational framework can be influenced by the choice of EOS, and more precisely, the choice of EOS parameter values in this case. Here, we present another simulation with a different group of EOS paramete values, listed as Group $2$ in Table~\ref{tab:EOS_params}. All the other (physical and numerical) parameters remain the same. 

\begin{figure}
  \centerline{\includegraphics[width = 0.95\textwidth]{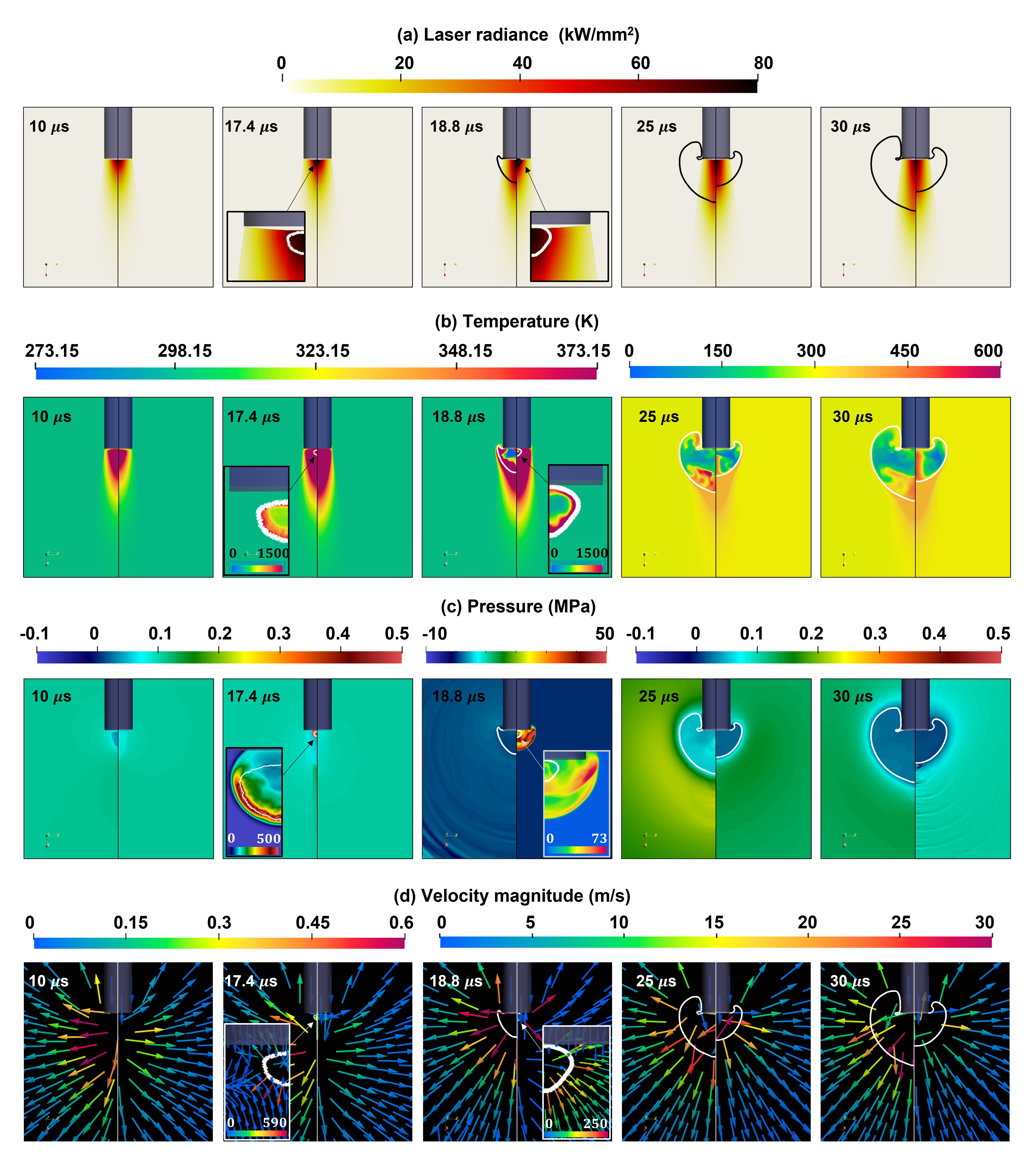}}
  \caption{Vapor bubble generated by a Ho:YAG laser: Side-by-side comparison of simulation results obtained with different EOS parameter values. In each sub-figure, the left and right halves show, respectively, the result obtained with Group $1$ and $2$ parameter values.}
 \label{fig:HoYAG_EOS}
\end{figure}

Figure~\ref{fig:HoYAG_EOS} compares the solutions obtained from the two simulations at five time instants. In each sub-figure, the left half of the figure is the solution obtained with Group $1$ parameter values, and the right half is the solution obtained with Group $2$ parameter values. Overall, both simulations predict the nucleation of a vapor bubble over a very short period of time. After that, the high pressure inside the bubble drives its expansion. In both simulations, a rounded bubble is obtained at the end.
 
However, a few differences can be found between the two simulations. First, there is a difference in the time when vaporization occurs. With Group $1$ parameter values, vaporization takes place at $17.4~\text{\textmu s}$. With Group $2$ parameter values, it occurs later, at $18.8~\text{\textmu s}$. The laser parameters are the same in both cases. Therefore, this discrepancy should be attributed mainly to the different temperature increase rates determined by the EOS, as defined in Eq.~\eqref{eq:Temperature_law}. Moreover, the speed of bubble growth is found to be lower with Group $2$ parameter values than with Group $1$. In Fig.~\ref{fig:HoYAG_EOS}, at both $25~\text{\textmu s}$ and $30~\text{\textmu s}$, the bubble obtained with Group $1$ parameter values is clearly larger than that obtained with Group $2$. The difference in growth speed can be explained by the pressure field. As shown in Fig.~\ref{fig:HoYAG_EOS} (c), the bubble's internal pressure reaches a maximum of $73~\text{MPa}$ with Group $2$ parameter values. This is much lower than the maximum pressure obtained with Group $1$ parameter values, which is $500~\text{MPa}$. Furthermore, the difference in bubble size also influences the delivery of laser energy. The larger bubble obtained with Group $1$ parameter values allows the laser energy to be delivered further, as demonstrated in Fig.~\ref{fig:HoYAG_EOS}(a). Lastly, the shapes of the bubbles obtained from the two simulations are slightly different. A pear-shaped bubble is captured with Group $1$ parameter values. With Group $2$, the bubble is more rounded. This discrepancy arises because in the latter case, no significant velocity difference is found at the bubble surface, as shown in Fig.~\ref{fig:HoYAG_EOS}(d).

\section{An elongated bubble}
\label{sec:validation_TFL}

In this section, we investigate the generation of an elongated bubble using Thulium fiber laser (TFL). The experiment is conducted in the same acrylic water tank, but the laser wavelength, the beam geometry, and the power profile are different. The simulation setup is modified to match the new experiment. Because of these changes, the vapor bubbles obtained from the experiment and the simulation both have a long, conical shape, different from the pear-shaped bubble shown in Sec.~\ref{sec:validation_Ho}. Again, we examine the full-field solutions obtained from the simulation to explain the bubble and fluid dynamics. In addition, we also discuss the effect of bubble dynamics on the delivery of laser energy.

\subsection{Comparison of experimental and numerical results}
\label{sec:TFL_setup}

\subsubsection{Laboratory experiment}
\label{sec:TFL_experiment}

\begin{figure}
  \centerline{\includegraphics[width = 0.9\textwidth]{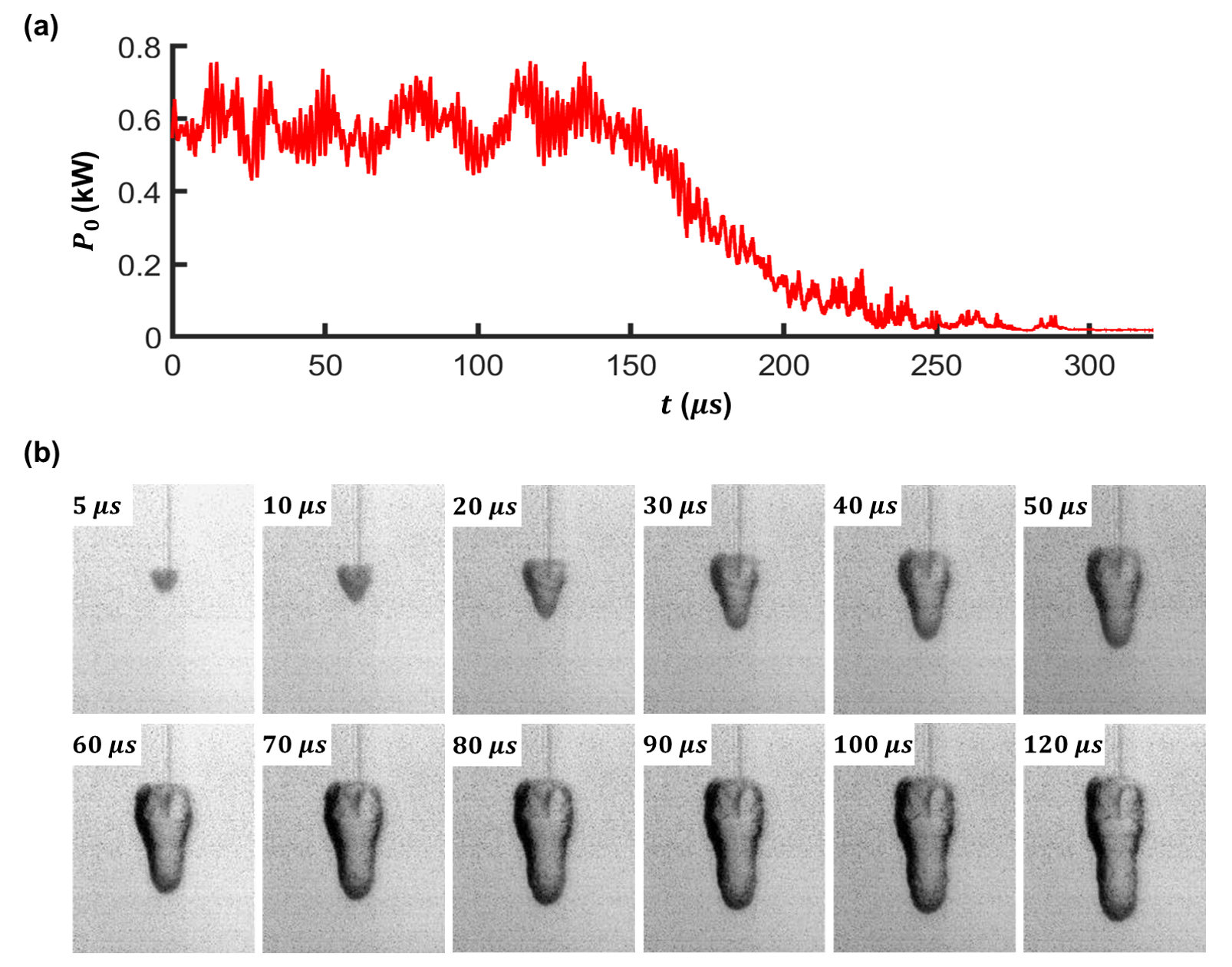}}
  \caption{The experimental result obtained with a Thulium fiber laser (TFL).
  (\textit{a}) Laser pulse profile measured in air, and (\textit{b}) high-speed images of the vapor bubble.}
 \label{fig:experiment_result_TFL}
\end{figure}

A Thulium Fiber Laser (TFL-$50$/$500$-QCW-AC, IPG Photonics, Oxford, MA) with a $1940$ nm wavelength is used to generate the vapor bubble. The laser generator is operated at the energy level of $0.11~\text{J}$, which is roughly half of the pulse energy in the Ho:YAG experiment (Sec.~\ref{sec:HoYAG_experiment}). The diameter of the laser fiber remains the same, but the laser beam is narrower \citep{blackmon2010holmium}. Again, the time-history of laser power is measured in air using a light detector and an oscilloscope (Fig.~\ref{fig:experiment_setup_Ho}(b)). Figure~\ref{fig:experiment_result_TFL}(a) shows the obtained result. The power profile exhibits a trapezoidal shape, with the laser power fluctuating around $0.6~\text{kW}$ for the first $140~\text{\textmu s}$. Then, it gradually decays to zero. Compared to the Ho:YAG laser (Fig.~\ref{fig:experiment_result_Ho}(a)), the pulse duration is longer, but the peak power is lower.

Figure~\ref{fig:experiment_result_TFL}(b) shows $12$ high-speed images of the vapor bubble during its nucleation and expansion stage ($0-120 ~\text{\textmu s}$). It can be observed that the laser pulse generates an elongated vapor bubble, clearly different from the pear-shaped bubble obtained with the Ho:YAG laser. Starting from the first frame at $t = 5~\text{\textmu s}$, a small bubble appears at the fiber tip. This means vaporization starts at a time between $0$ and $5~\text{\textmu s}$, earlier than that in the Ho:YAG experiment. From $5~\text{\textmu s}$ to $150~\text{\textmu s}$, the bubble grows continuously, forming a long, conical shape.

\subsubsection{Numerical simulation}
\label{sec:TFL_simulation}

We simulate the TFL experiment using the computational framework described in Secs.~\ref{sec:physical_model} and~\ref{sec:comp_framework}. The same computational domain and mesh described in Sec.~\ref{sec:HoYAG_simulation} are adopted. The geometry of the embedded laser fiber is also the same. The divergence angle of the laser beam, $\theta_{\text{div}}$, is set to $9.78^\circ$. This is because the wavelength of TFL is different from that of the Ho:YAG laser. The absorption coefficient, $\mu_\alpha$, is set to $14~\text{mm}^{-1}$ in liquid water \citep{traxer2020thulium}, which is about $6$ times the value for Ho:YAG laser. This is also due to the fact that TFL has a different wavelength. The absorption coefficient in water vapor is set to $0.001~\text{mm}^{-1}$, same as in the previous case.

\begin{figure}
  \centerline{\includegraphics[width = 0.95\textwidth]{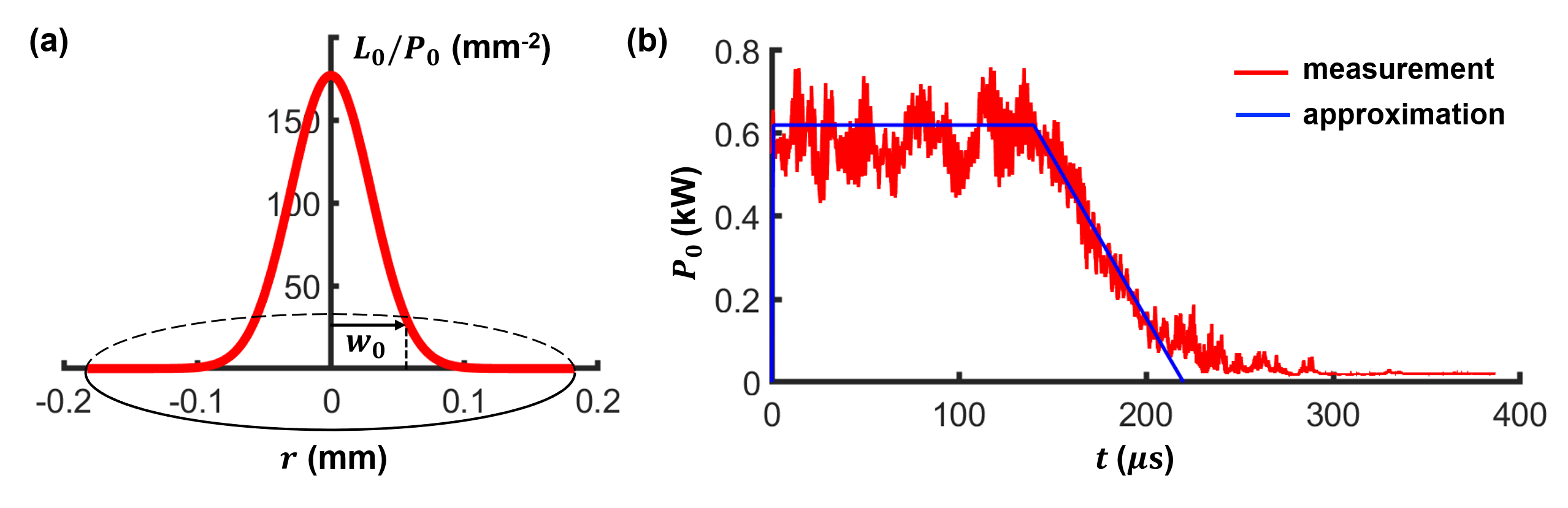}}
  \caption{Vapor bubble generated by a TFL: Simulation setup.
  (\textit{a}) Spatial profile of laser radiance on the source plane.  (\textit{b}) Temporal profile of laser power.}
 \label{fig:numerical_setup_TFL}
\end{figure}

The waist radius of the Gaussian beam is set to $0.05~\text{mm}$ (Fig.~\ref{fig:numerical_setup_TFL}(a)). The temporal profile of the laser power is specified to be a trapezoidal function that approximates the experimental measurement (Fig.~\ref{fig:numerical_setup_TFL}(b)). The resulting pulse energy is the same as that in the experiment, i.e., $0.11~\text{J}$. More specifically, the power grows rapidly from $0$ to $0.62~\text{kW}$ within $0.1~\text{\textmu s}$. This peak power is maintained for a period of $140~\text{\textmu s}$. Then, it vanishes gradually within $80~\text{\textmu s}$. The parameters of the Noble-Abel stiffened gas EOS are set by the values in Group $1$ in Table~\ref{tab:EOS_params}.

\begin{figure}
  \centerline{\includegraphics[width = 0.98\textwidth]{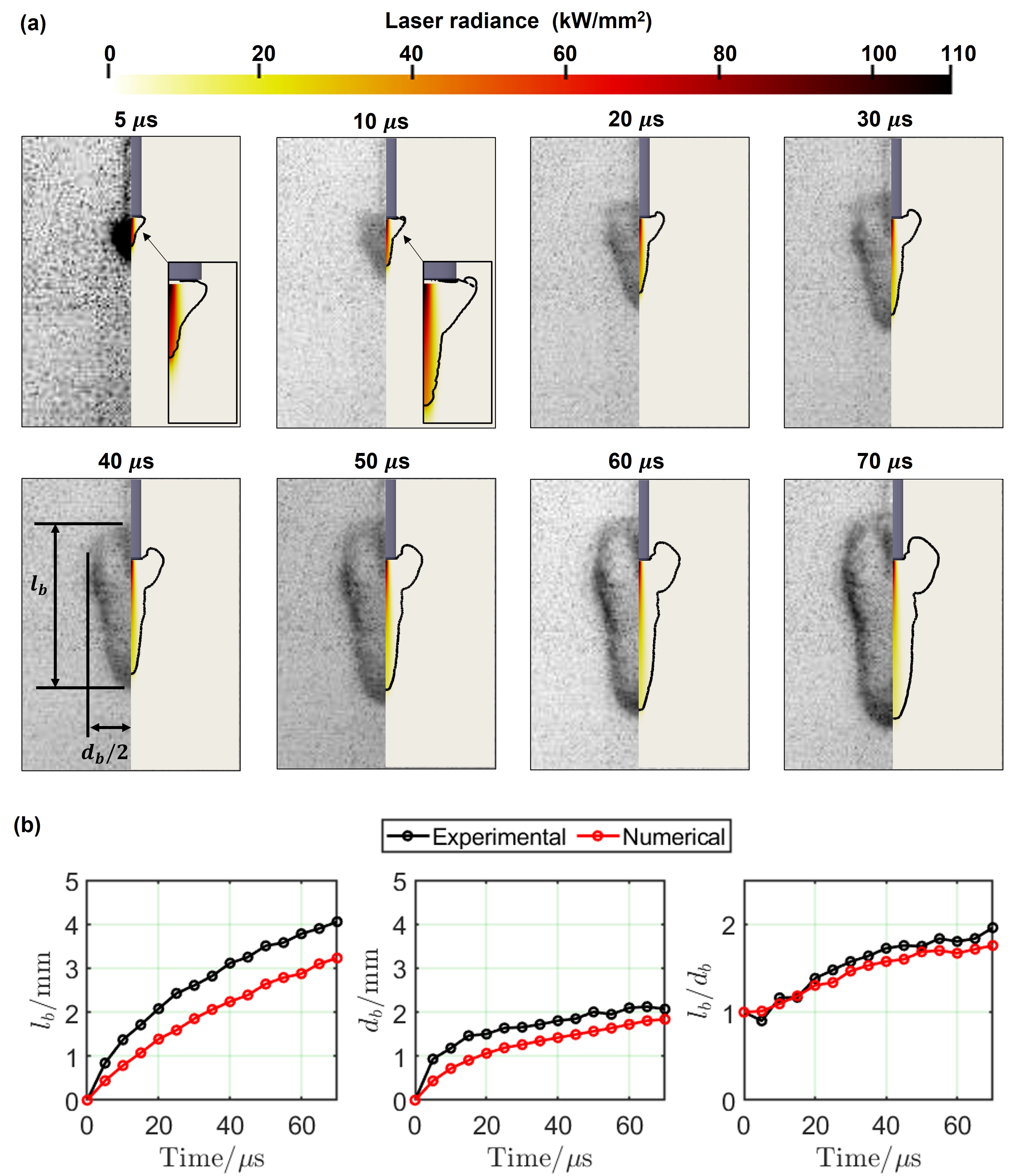}}
  \caption{Vapor bubble generated by a TFL: Comparison of bubble dynamics obtained from numerical simulation and laboratory experiment. (\textit{a}) Bubble nucleation and evolution. In each sub-figure, the left side shows the imaging result from the experiment, and the right side shows the bubble and laser radiance field predicted by the simulation. (\textit{b}) Evolution of bubble size and shape. $l_b$ and $d_b$ denote the maximum length of the bubble along and perpendicular to the laser fiber direction, respectively.}
 \label{fig:comp_TFL}
\end{figure}

The simulation predicts the formation of an elongated bubble, similar to the one observed in the experiment. Figure~\ref{fig:comp_TFL}(a) presents a side-by-side comparison between the results obtained from the experiment and the simulation. In each sub-figure, the left side is a high-speed image obtained from the experiment. The right side is the simulation result at the same time, showing the bubble surface and the laser radiance field. It can be seen that the bubble obtained from the simulation also has a long, conical shape. In the simulation, vaporization starts at $1.2~\text{\textmu s}$. This is consistent with the experimental data, which shows the bubble nucleates at a time between $0$ and $5~\text{\textmu s}$. At $5~\text{\textmu s}$, the shape of the bubble obtained from the simulation matches the experimental data reasonably well. As time progresses, the bubble from the simulation undertakes the same evolution trend, growing faster in the axial direction than in the radial direction. The main difference between the simulation result and the experimental data lies in the size of the bubble. The simulated bubble is smaller than its experimental counterpart in all the snapshots shown in Fig.~\ref{fig:comp_TFL}(a).

To make a quantitative comparison, we measure the length and width of the bubble, denoted by $l_b$ and $d_b$, respectively. Figure~\ref{fig:comp_TFL}(b) shows the time-histories of $l_b$, $d_b$, and the aspect ratio, $l_b/d_b$. It can be observed that the aspect ratio predicted by the simulation matches well the experimental result. For $l_b$ and $d_b$, the simulation is able to capture the same trend found in the experiment. But the magnitude is lower. It is notable that in both the simulation and the experiment, the aspect ratio starts at approximately $1$. Then, it increases steadily, reaching around $2$ after $70~\text{\textmu s}$. This implies that the bubble elongates gradually. In comparison, in the previous case with the Ho:YAG laser, the $l_b$-to-$d_b$ aspect ratio is roughly constant in time (Fig.~\ref{fig:comp_HoYAG}(b)). 

\subsection{Bubble elongation due to continuous vaporization}
\label{sec:TFL_bubble}

To investigate the mechanism of bubble elongation, we examine the temperature, pressure, and velocity fields obtained from the simulation. 

\begin{figure}
  \centerline{\includegraphics[width = 0.98\textwidth]{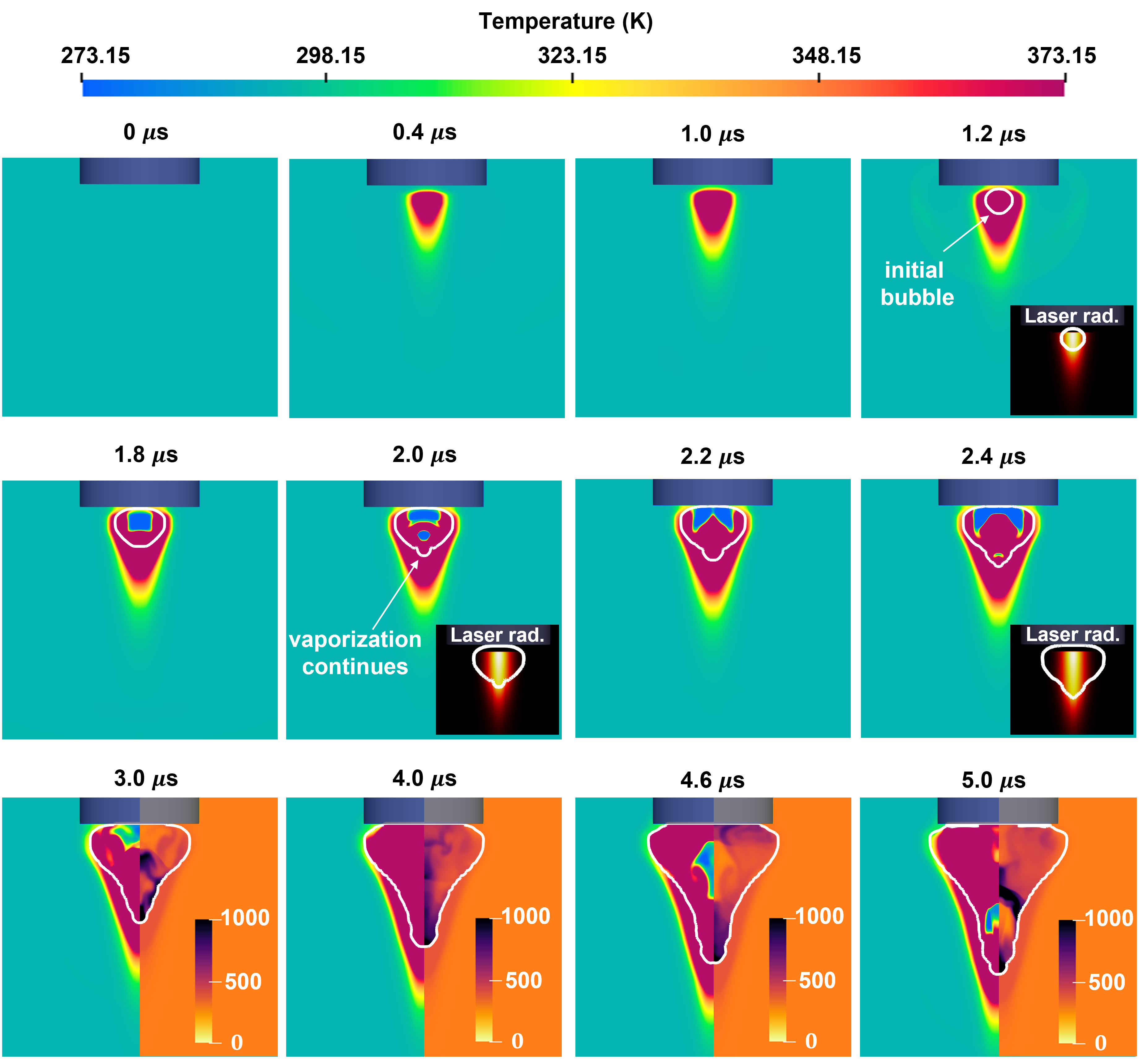}}
  \caption{Vapor bubble generated by a TFL: Evolution of the temperature field in the first $5~\text{\textmu s}$. For the solutions between $3.0~\text{\textmu s}$ and $5.0~\text{\textmu s}$, a different color scheme and range is applied to clearly show the temperature variation inside the bubble. The solution of laser radiance is shown at $1.2$, $2.0$, and $2.4~\text{\textmu s}$ (color range: $0-110~\text{kW}/\text{mm}^2$) as a reference.}
 \label{fig:temperature_TFL}
\end{figure}

Figure~\ref{fig:temperature_TFL} presents the evolution of the temperature field near the fiber tip in the first $5~\text{\textmu s}$. The laser radiance field is also shown at three time instants ($1.2$, $2.0$, and $2.4~\text{\textmu s}$) to facilitate the discussion. Similar to the previous case (Fig.~\ref{fig:temperature_HoYAG}), water temperature increases due to the absorption of laser energy, especially in the region around the central axis of the laser beam. The time it takes to reach the vaporization temperature is significantly shorter, only $0.4~\text{\textmu s}$, compared to $7~\text{\textmu s}$ in the previous case. The faster temperature increase can be attributed to two factors. First, the smaller beam waist of the laser results in a more concentrated distribution of laser radiance on the source plane. Although the laser power is lower in the current case (compare Figs.~\ref{fig:comp_HoYAG}(d) and~\ref{fig:comp_TFL}(b)), the more concentrated distribution leads to a higher laser radiance along the central axis, that is, $110~\text{kW/mm}^2$ (Fig.~\ref{fig:comp_TFL}(a)) versus $80~\text{kW/mm}^2$ in the previous case (Fig.~\ref{fig:comp_HoYAG}(a)). Second, the absorption coefficient of TFL in water is significantly higher than that of the Ho:YAG laser used previously ($14~\text{mm}^{-1}$ versus $2.42~\text{mm}^{-1}$). The time delay in bubble nucleation is also observed in this case. After $T_{\text{vap}}$ is reached, it takes another $0.8~\text{\textmu s}$ before vaporization occurs at the fiber tip, at $1.2~\text{\textmu s}$. Within this time period, the absorbed laser energy is converted to the intermolecular potential energy of liquid water.

A major difference from the previous case is that with the TFL, vaporization continues for a much longer period of time, that is, from $1.2~\text{\textmu s}$ until $53.5~\text{\textmu s}$. Also, it happens mainly along the central axis of the laser beam, which drives the bubble to grow in the same direction. 

Initially, a small, rounded bubble emerges in front of the laser fiber. This is shown in Fig.~\ref{fig:temperature_TFL}, in the snapshot taken at $1.2~\text{\textmu s}$. Because the vapor phase does not absorb laser energy ($\mu_\alpha$ set to $0.001~\text{mm}^{-1}$), the small bubble extends the laser beam along the axial direction. This effect can be seen in the inset images in Fig.~\ref{fig:temperature_TFL}. As a result, the liquid water next to the bubble's forward end --- that is, the forward-most point along the axial direction --- experiences a sudden increase in laser radiance, which accelerates the accumulation of energy there to overcome the latent heat of vaporization. From the simulation result, we observe the continuation of phase transition in this direction. For example, the snapshot taken at $2.0~\text{\textmu s}$ captures a bulge at the bubble's forward end, which is a newly vaporized region. At the same time, the high pressure inside the bubble also drives it to expand in both axial and radial directions. The combination of these two processes --- that is, phase transition and advection --- drives the bubble to grow into a long, conical shape. 

\begin{figure}
  \centerline{\includegraphics[width = 0.98\textwidth]{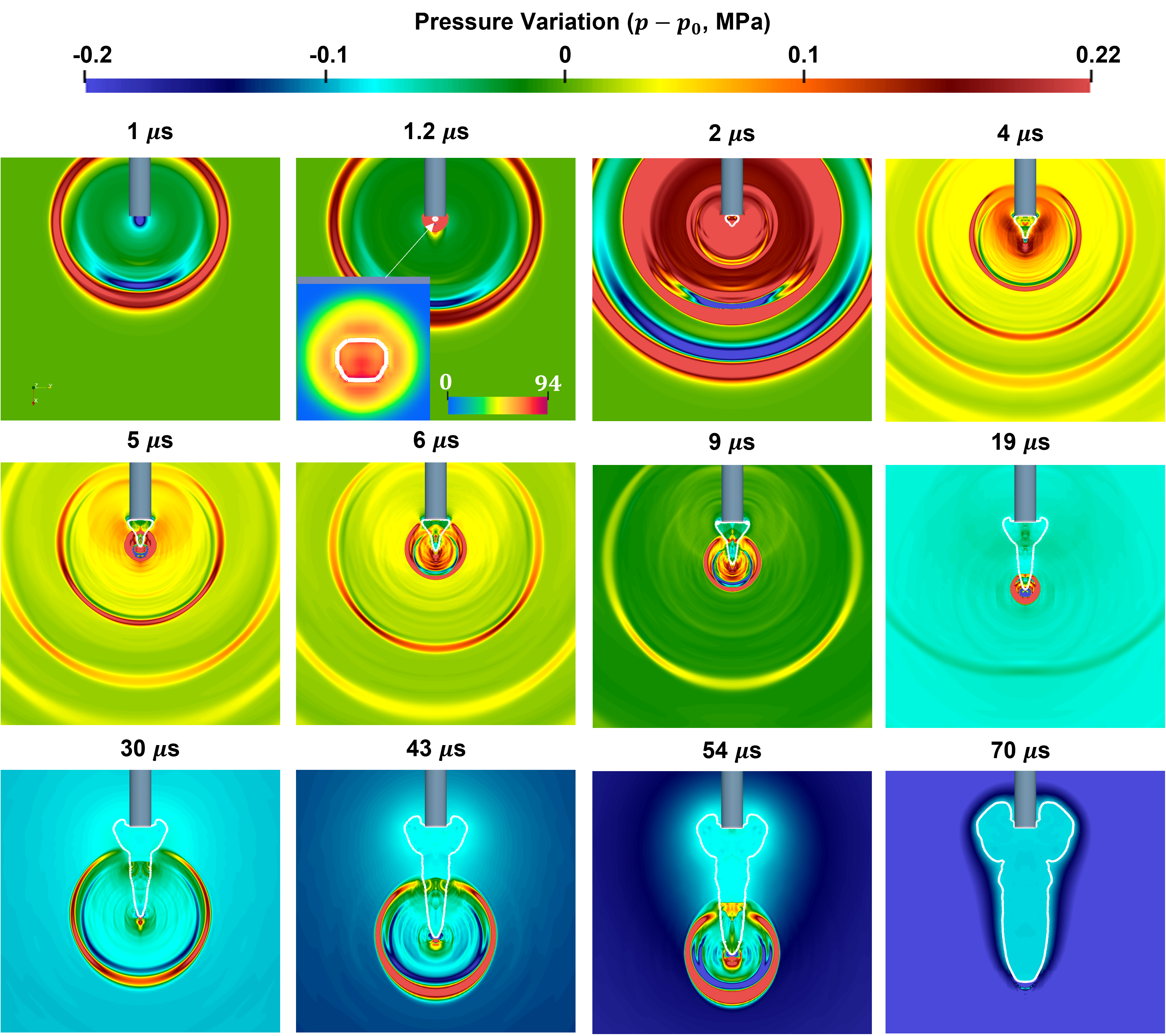}}
  \caption{Vapor bubble generated by a TFL: Evolution of the pressure field.}
 \label{fig:pressure_TFL}
\end{figure}

Figure~\ref{fig:pressure_TFL} presents the evolution of the pressure field up to $70~\text{\textmu s}$. At the beginning, the sudden increase of temperature due to the absorption of laser leads to a thermal shock at the fiber tip. The snapshot taken at $1~\text{\textmu s}$ captures this phenomenon, where the maximum pressure is found to be less than $0.5~\text{MPa}$. Next, the snapshot taken at $1.2~\text{\textmu s}$ captures the pressure field inside and around the initial bubble. The peak pressure inside the bubble is found to be $94~\text{MPa}$ at this time. This high pressure drives the bubble to expand in all directions. It also generates a weak shock wave that propagates outwards, at approximately the speed of sound in liquid water. Because of the small size of the initial bubble, the pressure magnitude of this wave quickly decays to less than $1$ MPa. Compared to the pressure field obtained from the Ho:YAG laser (Fig.~\ref{fig:pressure_HoYAG}), the main difference here is that as phase transition continues, acoustic waves keep emanating from the bubble's forward tip. This phenomenon is captured by all the snapshots taken between $2~\text{\textmu s}$ and $50~\text{\textmu s}$. In the current simulation, phase transition stops at $53.5~\text{\textmu s}$. Afterward, the pressure field becomes quiet. As shown in the snapshot at $70~\text{\textmu s}$, the main feature is that the bubble's internal pressure is higher than the ambient value. Therefore, the bubble continues growing. By this time, it has already formed a long, conical shape.

\begin{figure}
  \centerline{\includegraphics[width = 0.98\textwidth]{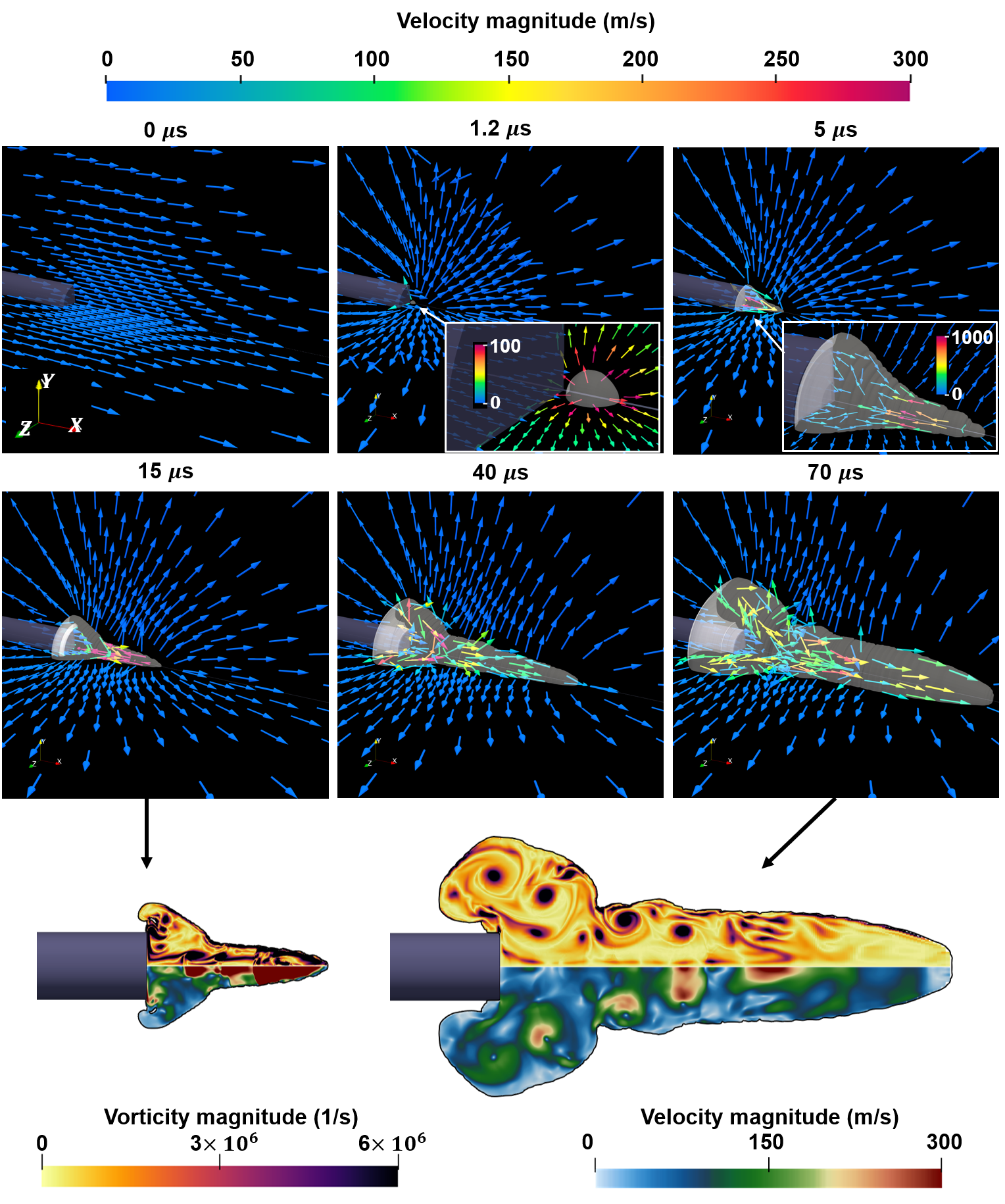}}
  \caption{Vapor bubble generated by a TFL: Evolution of the velocity field. The solution fields of velocity and vorticity magnitude inside the vapor bubble are shown for the time instances, $15~\text{\textmu s}$ and $70~\text{\textmu s}$, respectively.}
 \label{fig:velocity_TFL}
\end{figure}

Figure~\ref{fig:velocity_TFL} shows the evolution of the velocity field. The inset image at $1.2~\text{\textmu s}$ shows that when the initial bubble has just formed, the high internal pressure leads to high velocity in both axial and radial directions. This phenomenon is also found in the previous case (Fig.~\ref{fig:velocity_HoYAG}, $17.4~\text{\textmu s}$). In the previous case, the bubble's velocity quickly starts to decrease. In the current case, however, the velocity inside the bubble remains high. This is again because of the continuation of phase transition, as it keeps adding energy to the existing bubble. In addition, multiple vortexes are observed inside the vapor bubble as shown in the snapshots at $70~\text{\textmu s}$, which is related to the propagation of multiple acoustic waves emitted from the bubble's forward tip. Therefore, the evolution of the velocity field also suggests that phase transition plays a substantial role in the bubble's dynamics.

\subsection{Moses effect}

\begin{figure}
  \centerline{\includegraphics[width = 0.98\textwidth]{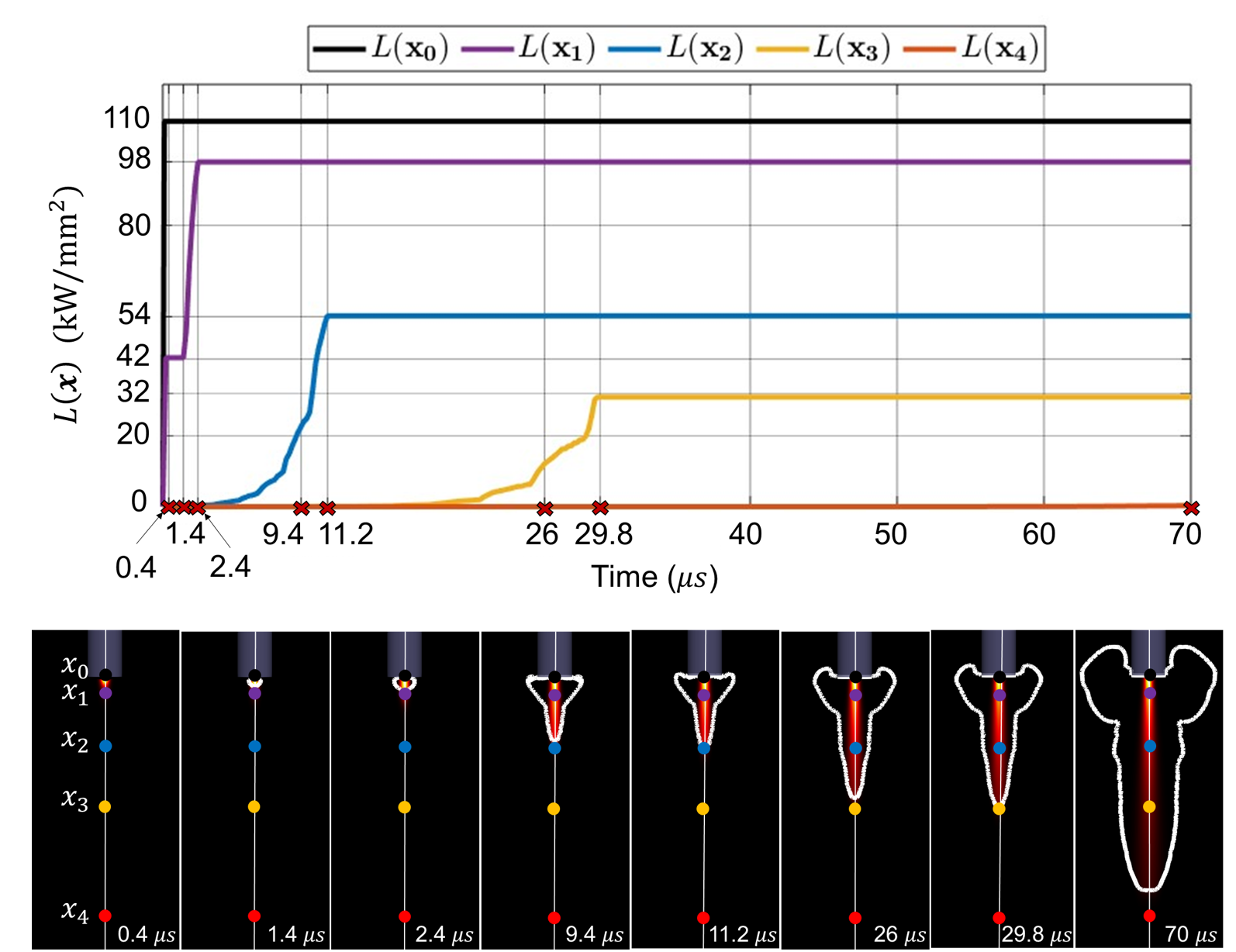}}
  \caption{Vapor bubble generated by a TFL: Moses effect.}
 \label{fig:moseseffect_TFL}
\end{figure}

Compared to liquid water, the absorption of laser by water vapor is negligible. Therefore, the formation of a vapor bubble along the path of the laser beam allows laser energy to be transmitted over a longer distance. This phenomenon, shown in Fig.~\ref{fig:temperature_TFL}, is sometimes referred to as Moses effect, after the story of Moses parting the sea \citep{ventimiglia2019moses,van1993bubble}. To investigate this effect more closely, we introduce four sensor points along the central axis of the laser beam, at difference distances from the fiber tip. Their coordinates (in mm) are $\bm{x}_1:(-0.3,0,0)$, $\bm{x}_2:(0.3,0,0)$, $\bm{x}_3:(1.1,0,0)$, and $\bm{x}_4:(2.7,0,0)$. The fiber tip is at $\bm{x}_0:(-0.5,0,0)$, as shown in Fig.~\ref{fig:numerical_setup_HoYAG}(b). Figure~\ref{fig:moseseffect_TFL} shows the time-history of laser radiance at the four sensor locations, as well as the fiber tip. 

Before bubble nucleation ($1.2~\text{\textmu s}$), most of the laser energy is absorbed by a small volume of water next to the fiber tip. Although Sensor $1$ is only $0.2~\text{mm}$ from the fiber tip, the laser radiance at this point has already dropped to $37.6\%$ of the value at the fiber tip. The laser radiance at the other sensor locations is negligible. This means without the vapor bubble, the laser cannot reach Sensors $2$, $3$, and $4$. 

After bubble nucleation, the laser radiance at all the sensor points starts to increase. At $2.4~\text{\textmu s}$, the bubble reaches $\bm{x}_1$. At this time, the laser radiance at Sensor $1$ reaches the maximum value, $98~\text{kW}/\text{mm}^2$. It is still lower than the laser radiance at the fiber tip, but this is only because the laser beam has a $9.78^\circ$ divergence angle. 

The time-histories of laser radiance at Sensors $2$ and $3$ follow the same trend. As the vapor bubble's forward end gets close to the sensor, laser radiance increases. The maximum value is reached when the bubble reaches the sensor. The maximum laser radiance decreases along the axial direction, due to beam divergence. Sensor $4$ is placed at $3.2~\text{mm}$ from the fiber tip. At $70~\text{\textmu s}$, the bubble's forward tip is still more than $0.29$ mm from it. As the result, the laser radiance at Sensor $4$ remains nearly zero.


In summary, the simulation result shows that the vapor bubble essentially creates a channel that allows laser to pass through. Compared to rounded bubbles, the long, conical shape obtained in the current case (both the experiment and the simulation) can be advantageous as it provides a longer channel.

\section{Transition between pear-shaped and elongated bubbles}
\label{sec:transition}

\subsection{A race between advection and phase transition}

Using different laser settings, we have obtained two vapor bubbles in different shapes, namely a rounded, pear-like shape shown in Sec.~\ref{sec:validation_Ho} and a long, conical shape shown in Sec.~\ref{sec:validation_TFL}. Depending on the application, one or the other may be preferred. By examining the simulation result, we find that a major difference between the two cases is that in the first case, vaporization only lasts for a short period of time, less than $1~\text{\textmu s}$. In the second case, vaporization continues along the laser beam direction for over $50~\text{\textmu s}$. It is also clear that in both cases, a newly vaporized region carries a high pressure (from the accumulated latent heat) that drives the existing bubble to expand by means of advection. Therefore, the simulation result suggests that when laser energy input is maintained in time (i.e.~long-pulsed laser), the vapor bubble's shape is influenced by two factors:
\begin{enumerate}
\item[(1)] the speed of bubble growth by advection, and
\item[(2)] the speed of bubble growth by phase transition. 
\end{enumerate}

Furthermore, the simulation result indicates that the transition between pear-shaped and elongated bubbles may be related to a competition between these two speeds. At least one of the two speeds must have changed from one case to the other. In other words, at least one of them is controllable.

Unfortunately, the fluid dynamics is highly nonlinear and multi-dimensional. It is impossible to separate the two speeds from the governing equations. In order to define and examine the two speeds analytically, we resort to a simplified model problem.

\begin{figure}
  \centerline{\includegraphics[width = 0.98\textwidth]{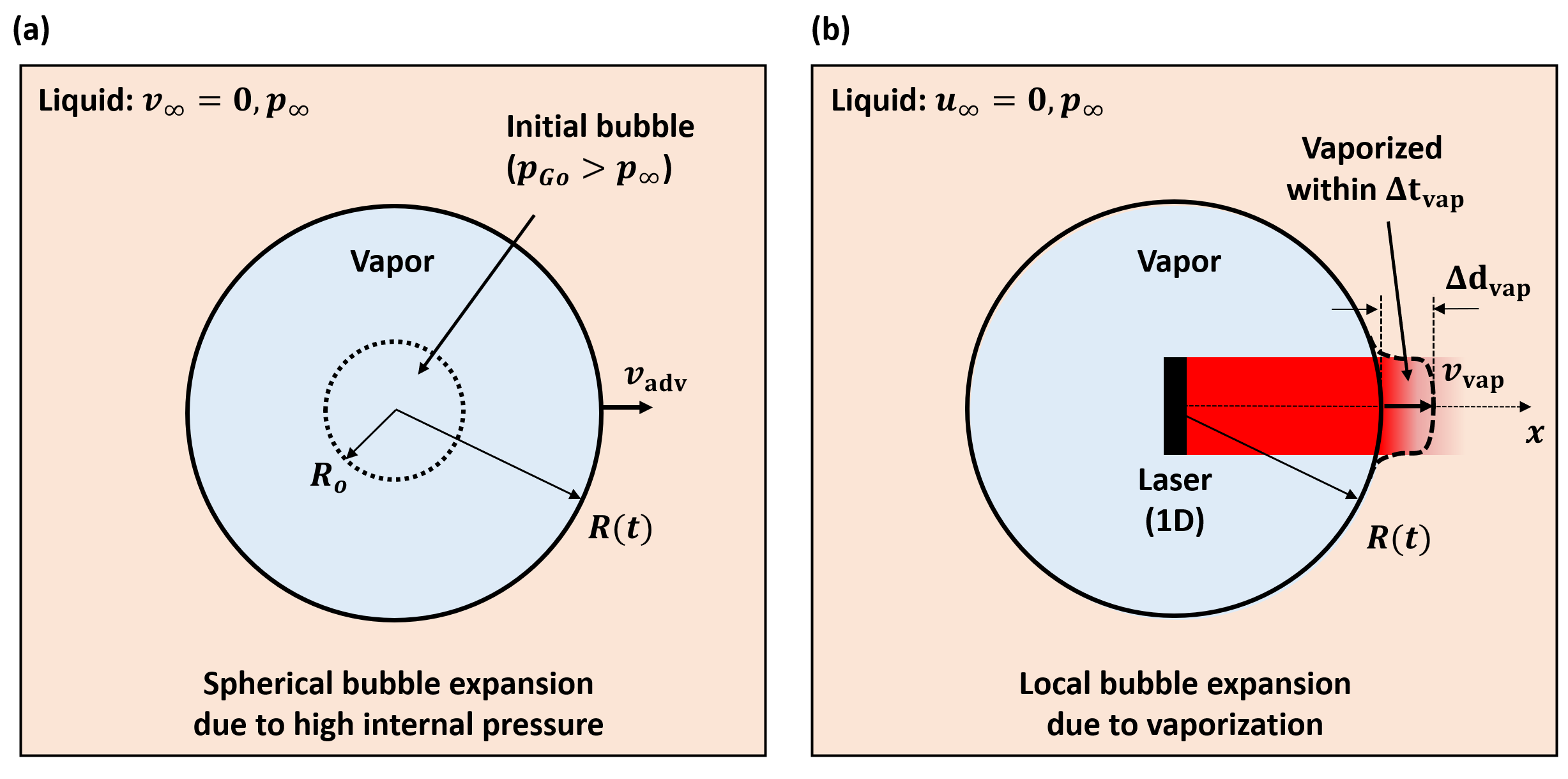}}
  \caption{Illustration of a simplified model problem for which the speeds of advection ($v_{\text{adv}}$) and phase transition ($v_{\text{vap}}$) are defined.}
 \label{fig:velocity_definition}
\end{figure}

As illustrated in Fig.~\ref{fig:velocity_definition}(a), we consider an initial vapor bubble of spherical shape, with radius $R_0$. We assume it has an internal pressure $p_{Go}$ that is higher than the ambient pressure $p_\infty$, which drives the bubble to expand. For this problem, the bubble dynamics can be modeled by the Rayleigh-Plesset equation \citep{brennen2014cavitation}. After dropping the viscosity and surface tension terms for simplicity, we get

\begin{equation}
    R \dfrac{d^2R}{dt^2} + \frac{3}{2} \left(\frac{dR}{dt}\right)^2 = \dfrac{p_{Go}}{\rho_0} \left[\left(\dfrac{R_0}{R}\right)^{3\gamma} - \dfrac{p_{\infty}}{p_{Go}}\right],
    \label{eq:Rayleigh-Plesset_1}
\end{equation}
where $\rho_0$ denotes the density of the liquid phase, and $\gamma$ the specific heat ratio of the vapor phase. In this case, the bubble's dynamics is isotropic, and driven only by advection. Therefore, we define the speed of bubble growth by advection as
\begin{equation}
    v_{\text{adv}}(t) = \dfrac{dR(t)}{dt},
    \label{eq:vadv_definition}
\end{equation}
where $R(t)$ is the solution of Eq.~\eqref{eq:Rayleigh-Plesset_1}. 

From Eq.~\eqref{eq:Rayleigh-Plesset_1}, it is clear that $v_{\text{adv}}$ depends on $p_{Go}$ and $R_0$. If the initial bubble is created through vaporization, $p_{Go}$ is determined by the thermodynamics of water, including its latent heat of vaporization (Sec.~\ref{sec:phase_transition}). Therefore, it may not always be adjustable. To see the effect of $R_0$, we first note that if we rewrite~\eqref{eq:Rayleigh-Plesset_1} with 
\begin{equation}
\tau = \dfrac{t}{R_0}\quad\text{and}\quad\widehat{R}(\tau)=\dfrac{R}{R_0},
\label{eq:RP_substitute}
\end{equation}
$R_0$ can be eliminated from the equation. Specifically, we have
\begin{equation}
    \widehat{R} \dfrac{d^2\widehat{R}}{d\tau^2} + \frac{3}{2} \left(\frac{d\widehat{R}}{d\tau}\right)^2 = \dfrac{p_{Go}}{\rho_L} \left[\left(\dfrac{1}{\widehat{R}}\right)^{3\gamma} - \dfrac{p_{\infty}}{p_{Go}}\right],
    \label{eq:Rayleigh-Plesset_2}
\end{equation}
and the solution $\widehat{R}(\tau)$ is independent of $R_0$. Also, substituting~\eqref{eq:RP_substitute} into~\eqref{eq:vadv_definition} yields
\begin{equation}
    v_{\text{adv}} = \dfrac{dR(t)}{dt} = \dfrac{d\widehat{R}(\tau)}{d\tau}.
    \label{eq:vadv_definition_2}
\end{equation}

Therefore, changing the value of $R_0$ leads to a linear scaling (i.e. stretching or compressing) of $v_{\text{adv}}$ in time, while the peak values remain the same. If the initial bubble is created by vaporization, $R_0$ may be controlled by adjusting the spatial distribution of the heat source. For example, a more uniform distribution of the laser power on the source plane may lead to a larger $R_0$.

Next, we assume that at a time $t>0$, a uniform, parallel laser beam is activated, and it creates a bulge on the bubble surface through vaporization (Fig.~\ref{fig:velocity_definition}(b)). We assume that over a short period of time, the vaporized region (i.e.~the bulge) has a cylindrical shape, with a depth of $\Delta d_{\text{vap}}$. To model the continuation of vaporization, we assume that at time $t$, $T = T_{\text{vap}}$ within this cylindrical region, and $\Lambda = l$ at $x=R(t)$. This assumption is justified by the results of the simulations shown in Secs.~\ref{sec:validation_Ho} and~\ref{sec:validation_TFL}.

The energy required to vaporize the forward end of the cylindrical region, i.e.~$x=R(t)+\Delta d_{\text{vap}}$, can be estimated by
\begin{equation}
\Delta E = \rho_0 \Big[l - \Lambda\big(R(t)+\Delta d_{\text{vap}},~t\big)\Big],
\label{eq:Delta_E}
\end{equation}
where $\rho_0$ denotes the density of the liquid phase, assumed to be a constant. The time to obtain this amount of energy from the laser beam can be estimated by
\begin{equation}
\Delta t_{\text{vap}} = \dfrac{\Delta E}{\mu_\alpha L\big(R(t)+\Delta d_{\text{vap}},~t\big)} = \rho_0 \dfrac{l - \Lambda\big(R(t)+\Delta d_{\text{vap}},~t\big)}{\mu_\alpha L\big(R(t)+\Delta d_{\text{vap}},~t\big)}.
\label{eq:vvap_dt}
\end{equation}

Now, we define the speed of bubble growth by phase transition as
\begin{equation}
    v_{\text{vap}} = \lim_{\Delta d_{\text{vap}} \to 0^+} \dfrac{\Delta d_{\text{vap}}}{\Delta t_{\text{vap}}}.
    \label{eq:vvap_def}
\end{equation}

Substituting~\eqref{eq:vvap_dt} into~\eqref{eq:vvap_def}, and noting that $\Lambda(R(t),t)=l$, we get
\begin{equation}
v_{\text{vap}}(t) = -\dfrac{\mu_\alpha L(R,t)}{\rho_0 \dfrac{\partial\Lambda}{\partial x}\Big|_{x=R(t)}}.
\label{eq:vvap_def_2}
\end{equation}

The derivative $\partial \Lambda/\partial x$ is negative, as the laser energy input decreases along the beam direction. Therefore, $v_{\text{vap}}$ is positive. It is notable that the latent heat of vaporization, $l$, is not involved in~\eqref{eq:vvap_def_2}. Intuitively, the latent heat represents an energy threshold for phase transition to occur. Once this threshold is reached, it may not influence the speed of bubble growth by phase transition. Eq.~\eqref{eq:vvap_def_2} also indicates that $v_{\text{vap}}$ may be controlled by adjusting the laser's wavelength and power, which will lead to variations in $\mu_\alpha$ and $L$.

\subsection{Testing hypothesis using simulation results}

We hypothesize that a race between advection and phase transition determines the morphing of the vapor bubble. In the previous subsection, we have defined speeds $v_{\text{adv}}$ and $v_{\text{vap}}$ for a simplified model problem. In real-world applications, these quantities would have to be estimated using available data. We hypothesize that at any time, if $v_{\text{adv}}$ is greater than $v_{\text{vap}}$, the bubble tends to grow spherically. If $v_{\text{adv}}$ is smaller than $v_{\text{vap}}$, the bubble tends to elongate along the laser beam direction.

\begin{figure}
  \centerline{\includegraphics[width = 0.7\textwidth]{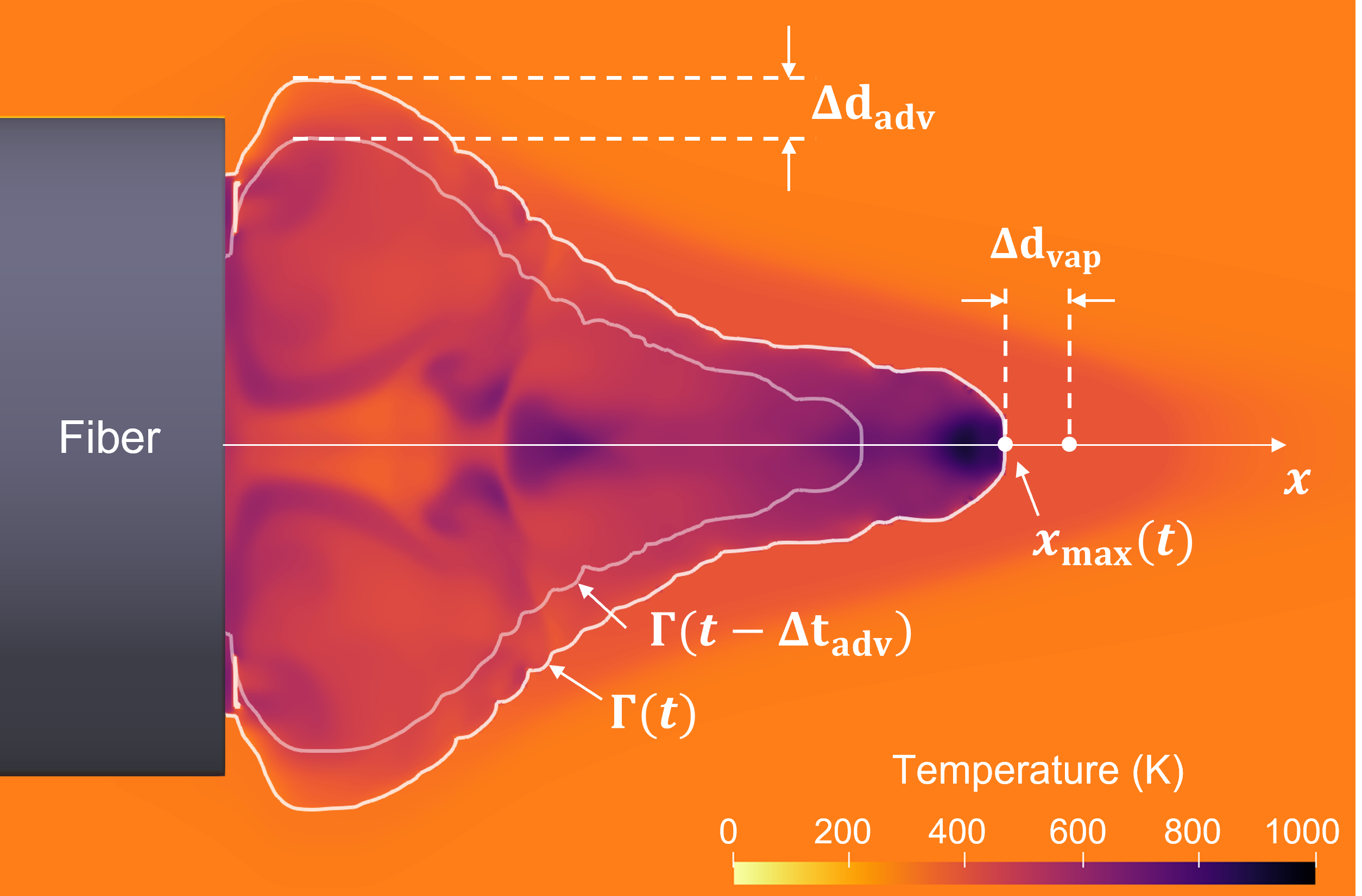}}
  \caption{Illustration of the method to estimate $v_{\text{adv}}$ and $v_{\text{vap}}$ using simulation results. The actual sizes of $\Delta d_{\text{adv}}$ and $\Delta d_{\text{vap}}$ used in the estimation are smaller than those shown in the figure. The temperature field is obtained from the simulation presented in Sec.~\ref{sec:validation_TFL}, at $t = 4.8~\text{\textmu s}$.}
 \label{fig:velocity_measurement}
\end{figure}

This hypothesis can be tested using our simulation results. Figure~\ref{fig:velocity_measurement} illustrates the method adopted here to estimate $v_{\text{adv}}$ and $v_{\text{vap}}$. Although this figure only shows a solution snapshot obtained from the TFL simulation, the same estimation method is also applied to the other simulation with a Ho:YAG laser. 

Because vaporization mainly continues along the laser beam direction, we estimate the advection speed, $v_{\text{adv}}$, by measuring the speed of bubble expansion in the radial direction, outside the beam waist. Specifically, we specify a small time interval, $\Delta t_{\text{adv}} = 0.2~\text{\textmu s}$. The radial expansion of the bubble over this time interval, $\Delta d_{\text{adv}}$ (Fig.~\ref{fig:velocity_measurement}), is measured at $255$ time points for the Ho:YAG simulation and $320$ time points for the TFL simulation. At each time point, $v_{\text{adv}}$ is estimated by
\begin{equation}
    \bar{v}_{\text{adv}} = \dfrac{\Delta d_{\text{adv}}}{\Delta t_{\text{adv}}}.
    \label{eq:vadv_measurement}
\end{equation}

To estimate the phase transition speed, $v_{\text{vap}}$, we look at the forward tip of the bubble, denoted by $x_{\text{max}}(t)$ in Fig.~\ref{fig:velocity_measurement}. We specify a small distance, $\Delta d_{\text{vap}} = 0.0015~\text{mm}$, along the laser beam direction. Then, we extract the simulation result of $L$ and $\Lambda$ to evaluate~\eqref{eq:vvap_dt}, which gives us $\Delta t_{\text{vap}}$, that is, an estimate of the time needed to extend the bubble front by $\Delta d_{\text{vap}}$. Then, we estimate $v_{\text{vap}}$ by
\begin{equation}
    \bar{v}_{\text{vap}} = \dfrac{\Delta d_{\text{vap}}}{\Delta t_{\text{vap}}},
    \label{eq:vvap_measurement}
\end{equation}
which is consistent with its definition in~\eqref{eq:vvap_def}. Again, we calculate  $\bar{v}_{\text{vap}}$ at $255$ time points for the Ho:YAG simulation and $320$ time points for the TFL simulation.

\begin{figure}
  \centerline{\includegraphics[width = 0.98\textwidth]{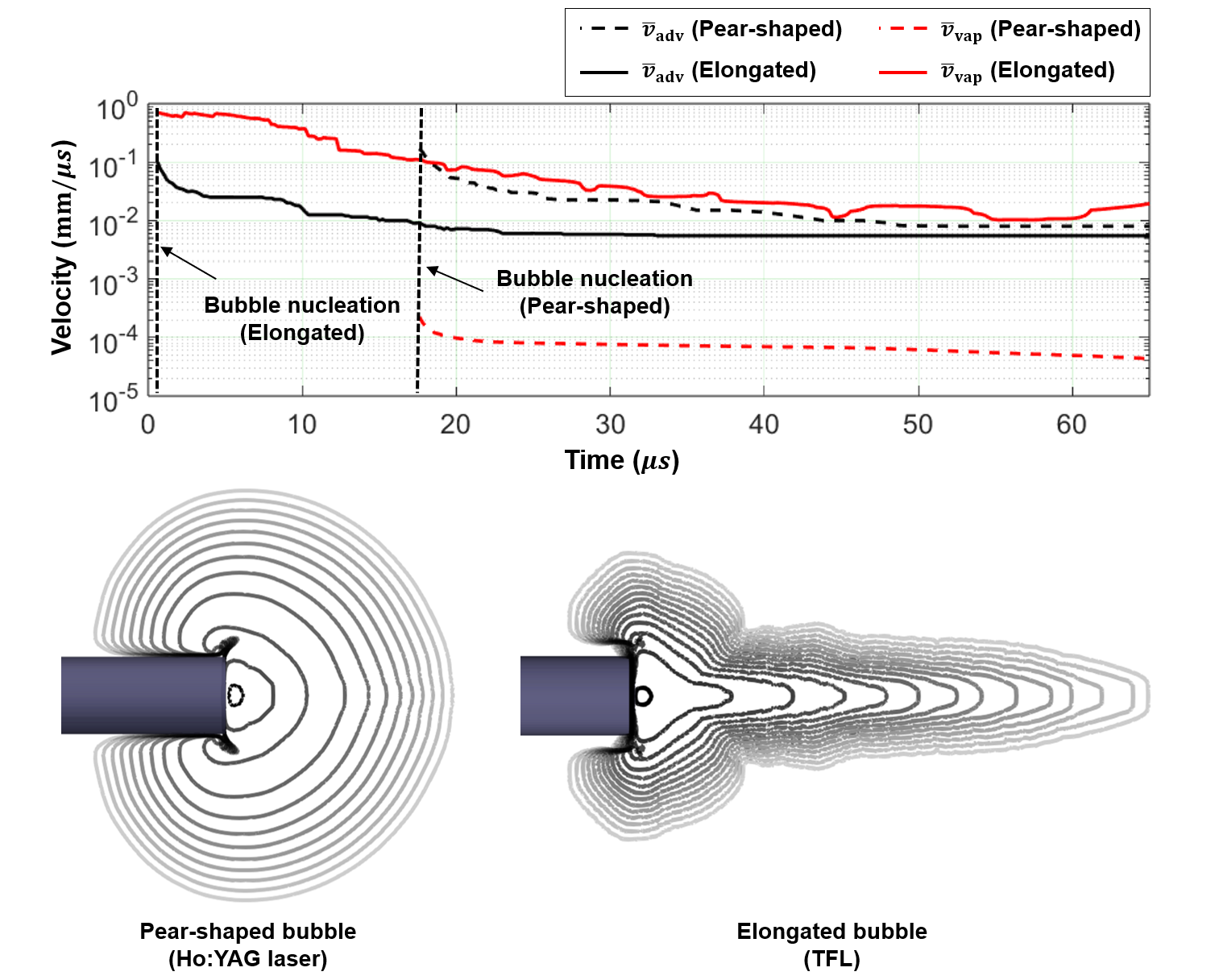}}
  \caption{Estimation of $v_{\text{adv}}$ and $v_{\text{vap}}$ for the pear-shaped bubble obtained with Ho:YAG laser (Sec.~\ref{sec:validation_Ho}) and the elongated bubbles obtained with TFL (Sec.~\ref{sec:validation_TFL}). The bubble dynamics is also shown by superimposing simulation results at different time instants.}
 \label{fig:velocity_comparison}
\end{figure}

Figure~\ref{fig:velocity_comparison} shows the time-histories of $\bar{v}_{\text{adv}}$ and $\bar{v}_{\text{vap}}$ obtained from the two simulations. For the Ho:YAG simulation that generated a pear-shaped bubble, $\bar{v}_{\text{adv}}$ is found to be roughly two orders of magnitude higher than $\bar{v}_{\text{vap}}$ over the entire time period shown in the figure. This is consistent with the finding that in this case, bubble expansion is mainly driven by advection, while phase transition only lasts for less than $1~\text{\textmu s}$. It also supports our hypothesis that if $v_{\text{adv}}$ is greater than $v_{\text{vap}}$, the bubble tends to grow spherically. For the TFL simulation that generated an elongated bubble, $\bar{v}_{\text{vap}}$ is found to be higher than $\bar{v}_{\text{adv}}$. Their difference is more than one order of magnitude in the early period of the simulation. But after $20~\text{\textmu s}$, the difference starts to become smaller. This is consistent with the finding that in this case, phase transition continues for a long period of time, until $53.5~\text{\textmu s}$. It also supports our hypothesis that if $v_{\text{adv}}$ is smaller than $v_{\text{vap}}$, the bubble tends to elongate along the laser beam direction.


In summary, the simulation results suggest that the transition between pear-shaped and elongated bubbles is determined by a race between advection and phase transition. These two speeds can be characterized by $v_{\text{adv}}$ and $v_{\text{vap}}$, which are mathematically defined for a simplified model problem. $v_{\text{adv}}$ and $v_{\text{vap}}$ depend on the laser setting and the properties of the fluid medium. Therefore, in real-world applications it may be possible to obtained a preferred bubble shape by adjusting the relevant parameters. For example, in our TFL experiment, the laser absorption coefficient and the source laser radiance are both higher than their counterparts in the Ho:YAG experiment. These changes lead to a significant increase of $v_{\text{vap}}$ by about $2$ orders of magnitude. In comparison, the variation of $v_{\text{adv}}$ is much smaller. Therefore, the changes in laser setting make $v_{\text{vap}}$ higher than $v_{\text{adv}}$. As a result, an elongated bubble is obtained.

\section{Concluding remarks}
\label{sec:conclusion}

In this work, we applied a laser-fluid computational model to study the physics behind vapor bubbles generated by long-pulsed lasers. The long pulse duration essentially means that three different physical processes --- laser radiation, phase transition (i.e.~vaporization), and fluid dynamics --- overlap both in time and in space. Their interaction adds complexity to the problem, but also makes it more interesting. Unlike short-pulsed lasers that usually produce spherical bubbles (assuming no influence from material boundaries), long-pulsed lasers can generate both rounded and elongated bubbles when operated in different settings.

In two separate laboratory experiments, we used Ho:YAG and Thulium fiber lasers to generate a rounded pear-shaped bubble and an elongated conical bubble. In each case, the laser power profile is also measured, and used as an input to the simulation. The computational model combines laser absorption, vaporization, and the dynamics and thermodynamics of a compressible two-phase fluid flow. To obtain a high resolution, the simulations are performed using a fine mesh that resolves the diameter of the laser fiber by more than $240$ elements. The two simulations for Ho:YAG and Thulium fiber lasers are performed with the same fluid parameters, such as the EOS parameters, the latent heat of vaporization, and thermal diffusivity. In both cases, the predicted bubble shape evolution matches the experimental data reasonably well. The simulation results show that the three physical processes mentioned above interact in multiple ways, including the following.
\begin{enumerate}
\item [(1)] The activation of laser radiation creates a thermal shock in the fluid flow.
\item [(2)] The absorption of laser increases the thermal energy and intermolecular potential energy of liquid water, eventually leading to its vaporization.
\item [(3)] The nucleation of a vapor bubble creates a new material subdomain (i.e.~vapor) in which laser can transmit almost losslessly.
\item [(4)] Because water has a high latent heat of vaporization, the bubble initially has a high internal pressure, which drives it to expand rapidly.
\item [(5)] The expansion of the bubble allows laser energy to be delivered over a greater distance. 
\end{enumerate}

Comparing the results of the two simulations, we find that the difference in bubble shape can be attributed to the duration of phase transition. In the case of the pear-shaped bubble, vaporization lasts for less than $1~\text{\textmu s}$. In the case of the elongated bubble, vaporization continues along the beam direction for over $50~\text{\textmu s}$. In both cases, the duration of the laser pulse is not a limiting factor. For example, the Ho:YAG laser that generated the pear-shaped bubble has a pulse duration of $70~\text{\textmu s}$, much longer than the time of vaporization.

The duration of phase transition can be explained as the result of a race between two bubble growth mechanisms, namely flow advection and the continuation of phase transition. The latter is a unique feature of long-pulse laser-induced cavitation. We hypothesize that at any time instant, if the speed of bubble growth by advection is higher than that by phase transition, the bubble tends to expand spherically. Otherwise, phase transition would occur (or continue), driving the bubble to elongate along the laser beam direction. We have formulated the two speeds using a simplified model problem, and estimated their values for the two experiments using the simulation results. The simulation results support the hypothesis. For example, the speed of bubble growth by phase transition is found to be two orders of magnitude higher in the case of the elongated bubble, while the speed of bubble growth by advection is about the same in the two cases. The formulas of bubble growth speeds also indicate possible ways to control the bubble shape, which can be a topic for future studies. For example, assuming the laser's power is fixed, increasing the laser absorption coefficient (e.g., by changing the laser's wavelength) and reducing the laser beam width may facilitate bubble elongation.


\backsection[Funding]{The authors gratefully acknowledge the support of the National Science Foundation (NSF) under Award CBET-1751487, the support of the Office of Naval Research (ONR) under Award N00014-19-1-2102, and the support of the National Institutes of Health (NIH) under Award 2R01-DK052985-24A1.}

\backsection[Declaration of interests]{The authors report no conflict of interest.}






\end{document}